%% file: 2026-SpkTracking-TwoCol.tex
\documentclass[journal]{IEEEtran}

\usepackage{algorithm}
\usepackage{algpseudocode}
\usepackage{amsmath,amsfonts}
\usepackage{array}
\usepackage[caption=false,font=normalsize,labelfont=sf,textfont=sf]{subfig}
\usepackage{textcomp}
\usepackage{stfloats}
\usepackage{url}
\usepackage{verbatim}
\usepackage{graphicx}
\usepackage[numbers,compress]{natbib}

\usepackage{microtype}

\usepackage{booktabs}
\usepackage{makecell}
\usepackage{multirow}
\usepackage{multicol}
\usepackage{tabularx}
\usepackage[dvipsnames]{xcolor}
\usepackage{caption}
\usepackage{shellesc}
\usepackage{tablefootnote} 
\usepackage{adjustbox}
\usepackage[flushleft]{threeparttable}
\usepackage{nomencl}
\makenomenclature

\usepackage[colorlinks=true, linkcolor=black, urlcolor=blue, citecolor=blue]{hyperref}
\usepackage{longtable}
\usepackage{supertabular}
\usepackage[symbols,acronym,automake]{glossaries-extra}

\newif\ifshowrevisions
\showrevisionsfalse
\input{Helpers/acronyms}

\input{Helpers/revision.tex}
\input{Helpers/symbols.tex}

\usepackage{orcidlink}

\newcommand{\etal}{\textit{et al.}}
\usepackage{gensymb}

\makeglossaries

\begin{document}

\title{{Speaker Tracking}: Segment-online Multi-talker Organization with a Varying Number of Speakers}

\author{
Vahid~A.~Kalkhorani\orcidlink{0000-0002-4166-1935}, Daniel~Wong\orcidlink{0000-0002-7781-1149}, Jacob~Donley\orcidlink{0000-0002-8401-798X}, Ashutosh~Pandey\orcidlink{0000-0002-3352-7453}, Buye~Xu\orcidlink{0000-0002-3027-7567}, and DeLiang~Wang\orcidlink{0000-0001-8195-6319}%
\thanks{
Vahid A. Kalkhorani (\href{ahmadikalkhorani.1@osu.edu}{ahmadikalkhorani.1@osu.edu}) is with the Department of Computer Science and Engineering, Ohio State University, Columbus, OH 43210 USA.
Daniel~Wong (\href{ddewong@meta.com}{ddewong@meta.com}), 
Jacob~Donley (\href{jdonley@meta.com}{jdonley@meta.com}), 
Ashutosh~Pandey (\href{apandey620@meta.com}{apandey620@meta.com}), 
and 
Buye Xu (\href{xub@meta.com}{xub@meta.com})
are with Meta Reality Labs, Redmond, WA, USA.
DeLiang Wang (\href{dwang@cuhk.edu.cn}{dwang@cuhk.edu.cn}) is with the School of Data Science, The Chinese University of Hong Kong, Shenzhen, China. All experiments were conducted at the Ohio State University.
}
}

\maketitle

\input{Sections/0-abstract}
\glsresetall

\input{Sections/01-Introduction}
\input{Sections/02-SystemOverview}

\input{Sections/03-Methodology}

\input{Sections/04-ExperimentalSetup}

\input{Sections/05-EvaluationResults}
\input{Sections/06-Conclusions}

\input{Sections/07-Acknowledgement}


\bibliographystyle{IEEEtran}

\bibliography{refs}



\end{document}

%% file: Helpers/acronyms.tex
\usepackage[acronym]{glossaries}
\setabbreviationstyle[acronym]{long-short}

\newacronym{ss}{SS}{speaker separation}
\newacronym{ssnd}{SSND}{speaker separation via neural diarization}
\newacronym{se}{SE}{speech enhancement}
\newacronym{tse}{TSE}{target speaker extraction}
\newacronym{sex}{SEx}{speaker extraction}
\newacronym{css}{CSS}{continuous speaker separation}
\newacronym{vad}{VAD}{voice activity detection}
\newacronym{eend}{EEND}{end-to-end neural diarization}
\newacronym{casa}{CASA}{computational auditory scene analysis}

\newacronym{ao}{AO}{audio-only}
\newacronym{csm}{CSM}{complex spectral mapping}
\newacronym{av}{AV}{audiovisual}
\newacronym{avse}{AVSE}{audiovisual speech enhancement}
\newacronym{avsr}{AVSR}{audiovisual speech recognition}
\newacronym{avss}{AVSS}{audiovisual speaker separation}
\newacronym{avtse}{AVTSE}{audiovisual target speaker extraction}

\newacronym{istft}{iSTFT}{inverse short-time Fourier transform}
\newacronym{stft}{STFT}{short-time Fourier transform}
\newacronym{tf}{T-F}{time-frequency}
\newacronym{rir}{RIR}{room impulse response}
\newacronym{rt60}{RT$_{60}$}{reverberation time}
\newacronym{mcmfwf}{MCMFWF}{multi-channel multi-frame Wiener filter}
\newacronym{scmfwf}{SCMFWF}{single-channel multi-frame Wiener filter}
\newacronym{asr}{ASR}{automatic speech recognition}
\newacronym{ri}{RI}{real and imaginary}

\newacronym{ffn}{FFN}{feed-forward networks}
\newacronym{deconv1d}{Deconv1D}{one-dimensional deconvolution}
\newacronym{mhsa}{MHSA}{multi-head self-attention}
\newacronym{gmhsa}{GMHSA}{global multi-head self-attention}
\newacronym{ln}{LN}{layer normalization}
\newacronym{bn}{BN}{batch normalization}
\newacronym{relu}{ReLU}{rectified linear unit}
\newacronym{prelu}{PReLU}{parametric rectified linear unit}
\newacronym{lnorm}{LN}{layer normalization}
\newacronym{glnorm}{gLN}{global layer normalization}
\newacronym{gnorm}{GN}{group normalization}
\newacronym{conv2d}{Conv2D}{2D convolutional layer}
\newacronym{conv1d}{Conv1D}{1D convolutional layer}
\newacronym{dnn}{DNN}{deep neural network}
\newacronym{cnn}{CNN}{convolutional neural network}
\newacronym{rnn}{RNN}{recurrent neural network}
\newacronym{lstm}{LSTM}{long short-term memory}
\newacronym{blstm}{BLSTM}{bidirectional long short-term memory}
\newacronym{gru}{GRU}{gated recurrent unit}
\newacronym{sgd}{SGD}{stochastic gradient descent}
\newacronym{adam}{Adam}{adaptive moment estimation}
\newacronym{bptt}{BPTT}{backpropagation through time}
\newacronym{nlp}{NLP}{natural language processing}
\newacronym{cv}{CV}{computer vision}
\newacronym{dm}{DM}{dynamic mixing}
\newacronym{pe}{PE}{positional encoding}
\newacronym{rpe}{RPE}{relative positional encoding}
\newacronym{rcpe}{RCPE}{random-chunk positional encoding}
\newacronym{tconvffn}{T-ConvFFN}{time-convolutional feed-forward network}
\newacronym{tconv}{T-Conv}{time-convolutional}
\newacronym{tgcon1d}{T-GConv1D}{grouped 1d convolution}
\newacronym{vtcn}{V-TCN}{visual temporal convolutional network}

\newacronym{silu}{SiLU}{sigmoid-weighted linear unit}
\newacronym{pit}{PIT}{permutation invariant training}
\newacronym{vbt}{VBT}{visual based training}
\newacronym{lbt}{LBT}{location-based training}
\newacronym{wm}{WM}{working memory}

\newacronym{nb-mhsa}{NB-MHSA}{narrow-band multi-head self-attention}

\newacronym{pesq}{PESQ}{perceptual evaluation of speech quality}
\newacronym{nb-pesq}{PESQ}{narrow-band perceptual evaluation of speech quality}
\newacronym{snr}{SNR}{signal-to-noise ratio}
\newacronym{tir}{TIR}{target-to-interferer ratio}
\newacronym{sdr}{SDR}{signal-to-distortion ratio}
\newacronym{sdri}{SDRi}{signal-to-distortion ratio improvement}
\newacronym{si-sdr}{SI-SDR}{scale-invariant signal-to-distortion ratio}
\newacronym{sa-sdr}{SA-SDR}{source-aggregated signal-to-distortion ratio}
\newacronym{si-sa-sdr}{SI-SA-SDR}{scale-invariant source-aggregated signal-to-distortion ratio}
\newacronym{si-snri}{SI-SNRi}{scale-invariant signal-to-noise ratio improvement}
\newacronym{si-sdrse}{SI-SDR(SE)}{scale-invariant signal-to-distortion ratio (scaled estimated)}
\newacronym{si-sdri}{SI-SDRi}{scale-invariant signal-to-distortion ratio improvement}
\newacronym{stoi}{STOI}{short-time objective intelligibility} 
\newacronym{estoi}{eSTOI}{extended short-time objective intelligibility} 
\newacronym{mc}{MC}{mixture constraint loss}
\newacronym{wer}{WER}{word error rate}
\newacronym{cpwer}{cpWER}{concatenated minimum-permutation word error rate }
\newacronym{der}{DER}{diarization error rate}
\newacronym{mi}{MI}{missed speech}
\newacronym{fa}{FA}{false alarm}
\newacronym{cf}{CF}{speaker confusion}

\newacronym{fps}{FPS}{frames per second}

\newacronym{flop}{FLOP}{floating point operation}
\newacronym{gflop}{GFLOP}{Giga floating point operation}

\newacronym{lrs}{LRS}{lip reading sentences}

\newacronym{fifo}{FIFO}{first-in-first-out}

\newacronym{ipd}{IPD}{inter-channel phase difference}
\newacronym{ild}{ILD}{inter-channel level difference}
\newacronym{itd}{ITD}{inter-channel time difference}
\newacronym{sad}{SAD}{speech activity detection}

%% file: Helpers/revision.tex
\usepackage{xcolor}
\usepackage{soul}
\usepackage{mdframed} 

\showrevisionsfalse 

\newlength{\docparindent}
\AtBeginDocument{\setlength{\docparindent}{\parindent}}

\soulregister\cite7
\soulregister\etal7
\soulregister\gls7
\soulregister\autoref7
\soulregister\ref7
\newcommand{\SIpct}[1]{\SI{#1}{\percent}}
\soulregister\SIpct7

\newcommand{\GetRevColor}[1]{%
  \ifcase#1\relax white\or orange!30\or blue!10\or violet!30\else white\fi
}

\makeatletter
\newcommand{\LogRevision}[2]{%
  \ifshowrevisions
      \def\temp{#1}\ifx\temp\empty\def\listtext{Revision (Level #2)}\else\def\listtext{#1}\fi
      \addcontentsline{lor}{revision}{\protect\numberline{Rev #2}\listtext}%
  \fi
}

\usepackage{colortbl} 

\newcommand{\revrow}[1]{%
  \ifshowrevisions\ifcase#1\relax\else\rowcolor{\GetRevColor{#1}}\fi\fi
}

\newcommand{\listofrevisions}{\ifshowrevisions\section*{List of changes}\@starttoc{lor}\fi}
\newcommand{\l@revision}{\@dottedtocline{1}{0em}{5.5em}}
\makeatother

\newcommand{\rev}[3][]{%
  \LogRevision{#1}{#2}%
  \ifshowrevisions
     \ifcase#2\relax#3\else\sethlcolor{\GetRevColor{#2}}\hl{#3}\fi
  \else
     #3
  \fi
}

\ifshowrevisions
    \newenvironment{revblock}[2][]{%
      \LogRevision{#1}{#2}%
      \begin{mdframed}[%
        backgroundcolor=\GetRevColor{#2},%
        hidealllines=true,%
        nobreak=false,
        skipabove=0pt,%
        skipbelow=0pt,%
        innerleftmargin=3pt,%
        innerrightmargin=3pt,%
        innertopmargin=3pt,%
        innerbottommargin=3pt,%
        leftmargin=-3pt,%
        rightmargin=-3pt%
      ]%
      \setlength{\parindent}{\docparindent}%
    }{%
      \end{mdframed}%
    }
\else
\fi

\ifshowrevisions
    \usepackage[colorinlistoftodos,color=yellow!20,textsize=scriptsize,linecolor=gray!40,textwidth=1.4in]{todonotes}
\else  
    \usepackage[disable]{todonotes}
\fi

%% file: Helpers/symbols.tex
\usepackage{physics}
\usepackage{amssymb}
\usepackage{pifont}
\usepackage{supertabular}
\usepackage{siunitx}
\usepackage[symbols,acronym,automake]{glossaries-extra}

\newglossarystyle{symbunitlong}{
    \setglossarystyle{super}
    
    \renewcommand*{\glsgroupheading}[1]{}

}

\newcommand*{\tran}{^{\mkern-1.5mu\mathsf{T}}}
\DeclareMathOperator*{\argmax}{arg\,max}
\DeclareMathOperator*{\argmin}{arg\,min}

\newcommand{\loss}[1]{\glssymbol{LossBase}_{\text{#1}}}

\newglossaryentry{nChn}{
    name={number of channels},
    symbol={M},
    description={number of audio channels},
    user1={$-$},
    sort={M},
    type=symbols
}
\newglossaryentry{iChn}{
    name={channel index},
    symbol={m},
    description={array channel index},
    user1={$-$},
    sort={m},
    type=symbols
}

\newglossaryentry{seqLen}{
    name={sequence length},
    symbol={L},
    description={length of the temporal sequence or overlapped region},
    user1={$-$},
    sort={L},
    type=symbols
}

\newglossaryentry{itime}{
    name={time index},
    symbol={t},
    description={discrete processing window or time frame index},
    user1={$-$},
    sort={t},
    type=symbols
}

\newglossaryentry{ifreq}{
    name={frequency index},
    symbol={f},
    description={frequency bin index},
    user1={$-$},
    sort={f},
    type=symbols
}

\newglossaryentry{nFreq}{
    name={total frequency bins},
    symbol={F},
    description={total number of frequency bins},
    user1={$-$},
    sort={F},
    type=symbols
}

\newglossaryentry{ispk}{
    name={speaker index},
    symbol={k},
    description={local target speaker index within a segment},
    user1={$-$},
    sort={k},
    type=symbols
}

\newglossaryentry{gspk}{
    name={global speaker index},
    symbol={n},
    description={global speaker index representing physical identity across time},
    user1={$-$},
    sort={n},
    type=symbols
}

\newglossaryentry{Gspk}{
    name={total global speakers},
    symbol={N},
    description={total number of distinct global speakers tracked up to a given time},
    user1={$-$},
    sort={N},
    type=symbols
}

\newglossaryentry{iseg}{
    name={segment index},
    symbol={\tau},
    description={temporal segment index},
    user1={$-$},
    sort={z_tau},
    type=symbols
}

\newglossaryentry{nEmb}{
    name={embedding dimension},
    symbol={D},
    description={dimensionality of the feature embeddings},
    user1={$-$},
    sort={D},
    type=symbols
}

\newglossaryentry{reals}{
    name={real numbers},
    symbol={\mathbb{R}},
    description={space of real numbers},
    user1={$-$},
    sort={R_bb},
    type=symbols
}

\newglossaryentry{identity}{
    name={identity matrix},
    symbol={\mathbb{I}},
    description={identity matrix},
    user1={$-$},
    sort={I_bb},
    type=symbols
}

\newglossaryentry{permSet}{
    name={permutation set},
    symbol={\mathcal{P}},
    description={set of all possible permutations of a given length},
    user1={$-$},
    sort={P_cal},
    type=symbols
}
\newglossaryentry{permSetWER}{
    name={permutation set for WER},
    symbol={\mathcal{Q}},
    description={set of all possible permutations of a given length for WER evaluation},
    user1={$-$},
    sort={Q_cal},
    type=symbols
}

\newglossaryentry{activeSet}{
    name={active speaker set},
    symbol={\mathcal{A}},
    description={set of locally active speaker indices in the overlap region of segment $\tau$, determined by $\mathcal{F}_{\text{VAD}}$; $\mathcal{A}_{\tau} \subseteq \{1,\ldots,K\}$},
    user1={$-$},
    sort={A_cal},
    type=symbols
}

\newglossaryentry{verificationThreshold}{
    name={verification threshold},
    symbol={\theta},
    description={decision threshold for speaker verification},
    user1={$-$},
    sort={z_theta},
    type=symbols
}

\newglossaryentry{stitchThreshold}{
    name={stitching threshold},
    symbol={\gamma},
    description={minimum \gls{si-sdr} required to accept an overlap-based stitching assignment},
    user1={\SI{dB}},
    sort={z_gamma},
    type=symbols
}

\newglossaryentry{cossim}{
    name={cosine similarity},
    symbol={\mathcal{C}},
    description={cosine similarity metric for embedding comparison},
    user1={$-$},
    sort={c},
    type=symbols
}

\newglossaryentry{distMetric}{
    name={distance metric},
    symbol={\mathcal{D}},
    description={distance metric for sequence grouping},
    user1={$-$},
    sort={D_cal},
    type=symbols
}

\newglossaryentry{optPerm}{
    name={optimal permutation},
    symbol={\pi^*},
    description={optimal alignment permutation for overlap-based stitching},
    user1={$-$},
    sort={z_pi_hat},
    type=symbols
}
\newglossaryentry{optSpk}{
    name={optimal speaker assignment index in working memory},
    symbol={\gspk^*},
    description={optimal global speaker assignment index in working memory for discontinuous grouping},
    user1={$-$},
    sort={n_start},
    type=symbols
}

\newglossaryentry{perm}{
    name={permutation},
    symbol={\pi},
    description={alignment permutation for overlap-based stitching},
    user1={$-$},
    sort={z_pi},
    type=symbols
}

\newglossaryentry{tsdreps}{
    name={epsilon},
    symbol={\varepsilon},
    description={small constant for SI-SDR numerical stability},
    user1={$-$},
    sort={z_epsilon_1},
    type=symbols
}

\newglossaryentry{sksdralpha}{
    name={SI-SDR alpha},
    symbol={\nu},
    description={scaling factor for SI-SDR},
    user1={$-$},
    sort={z_nu},
    type=symbols
}

\newglossaryentry{sisdralpha}{
    name={speaker SI-SDR alpha},
    symbol={\alpha},
    description={SI-SDR scaling factor for speaker $k$},
    user1={$-$},
    sort={z_alpha_k},
    type=symbols
}

\newglossaryentry{sisdrbeta}{
    name={SI-SDR beta},
    symbol={\beta},
    description={SI-SDR regularization parameter},
    user1={$-$},
    sort={z_beta},
    type=symbols
}

\newglossaryentry{sdrmax}{
    name={maximum SDR},
    symbol={\text{SDR}_{\text{max}}},
    description={maximum achievable signal-to-distortion ratio},
    user1={\SI{dB}},
    sort={S_sdr},
    type=symbols
}

\newglossaryentry{LossBase}{
    name={loss function},
    symbol={\mathcal{L}},
    description={base loss function},
    user1={$-$},
    sort={L_cal},
    type=symbols
}

\newglossaryentry{mixScaler}{
    name={mixture scalar},
    symbol={y},
    description={time-domain mixture signal scalar observation},
    user1={$-$},
    sort={y},
    type=symbols
}

\newglossaryentry{sourceScaler}{
    name={source scalar},
    symbol={s},
    description={time-domain target source signal scalar},
    user1={$-$},
    sort={s},
    type=symbols
}

\newglossaryentry{activityScaler}{
    name={activity scalar},
    symbol={a},
    description={time-domain speaker activity scalar},
    user1={$-$},
    sort={a},
    type=symbols
}

\newglossaryentry{noiseScaler}{
    name={noise scalar},
    symbol={n},
    description={time-domain noise signal scalar},
    user1={$-$},
    sort={n},
    type=symbols
}

\newglossaryentry{estimateScaler}{
    name={source estimate scalar},
    symbol={\hat{s}},
    description={estimated source stream scalar},
    user1={$-$},
    sort={s_hat},
    type=symbols
}

\newglossaryentry{estimateActivityScaler}{
    name={activity estimate scalar},
    symbol={\hat{a}},
    description={time-domain activity estimate scalar},
    user1={$-$},
    sort={a_hat},
    type=symbols
}

\newglossaryentry{mix}{
    name={mixture vector},
    symbol={\vb{y}},
    description={time-domain mixture signal segment},
    user1={$-$},
    sort={y_vec},
    type=symbols
}

\newglossaryentry{source}{
    name={source vector},
    symbol={\vb{s}},
    description={time-domain local target source vector for a segment},
    user1={$-$},
    sort={s_vec_k},
    type=symbols
}

\newglossaryentry{sourceOther}{
    name={other source vector},
    symbol={\vb{s}_{k^\prime}},
    description={time-domain source signal vector for interfering speaker},
    user1={$-$},
    sort={s_vec_k_prime},
    type=symbols
}

\newglossaryentry{noise}{
    name={noise vector},
    symbol={\vb{n}},
    description={time-domain background noise vector},
    user1={$-$},
    sort={n_vec},
    type=symbols
}

\newglossaryentry{estimate}{
    name={source estimate vector},
    symbol={\vb{\hat{s}}},
    description={time-domain local source estimate vector for a segment},
    user1={$-$},
    sort={s_vec_hat},
    type=symbols
}

\newglossaryentry{estimateActivity}{
    name={activity estimate vector},
    symbol={\vb{\hat{a}}},
    description={time-domain speaker activity estimate vector},
    user1={$-$},
    sort={a_vec_hat},
    type=symbols
}

\newglossaryentry{leadingSeg}{
    name={leading segment portion},
    symbol={\mathbf{p}},
    description={leading portion of the current acoustic segment},
    user1={$-$},
    sort={p},
    type=symbols
}

\newglossaryentry{trailingSeg}{
    name={trailing segment portion},
    symbol={\mathbf{q}},
    description={trailing portion of the previous acoustic segment},
    user1={$-$},
    sort={q},
    type=symbols
}

\newglossaryentry{mixSTFTScaler}{
    name={STFT mixture scalar},
    symbol={Y},
    description={STFT domain mixture scalar},
    user1={$-$},
    sort={Y},
    type=symbols
}

\newglossaryentry{sourceSTFTScaler}{
    name={STFT source scalar},
    symbol={S},
    description={STFT domain source scalar},
    user1={$-$},
    sort={S},
    type=symbols
}

\newglossaryentry{noiseSTFTScaler}{
    name={STFT noise scalar},
    symbol={N},
    description={STFT domain noise scalar},
    user1={$-$},
    sort={N},
    type=symbols
}

\newglossaryentry{estimateSTFTScaler}{
    name={STFT source estimate scalar},
    symbol={\hat{S}},
    description={STFT domain source estimate scalar},
    user1={$-$},
    sort={S_hat},
    type=symbols
}

\newglossaryentry{mixSTFT}{
    name={STFT mixture matrix},
    symbol={\vb{Y}},
    description={STFT domain mixture matrix},
    user1={$-$},
    sort={Y_vec},
    type=symbols
}

\newglossaryentry{sourceSTFT}{
    name={STFT source matrix},
    symbol={\vb{S}},
    description={STFT domain source matrix},
    user1={$-$},
    sort={S_vec},
    type=symbols
}

\newglossaryentry{noiseSTFT}{
    name={STFT noise matrix},
    symbol={\vb{N}},
    description={STFT domain noise matrix},
    user1={$-$},
    sort={N_vec},
    type=symbols
}

\newglossaryentry{estimateSTFT}{
    name={STFT source estimate matrix},
    symbol={\vb{\hat{S}}},
    description={STFT domain source estimate matrix},
    user1={$-$},
    sort={S_vec_hat},
    type=symbols
}

\newglossaryentry{video}{
    name={video features},
    symbol={\vb{V}},
    description={extracted video feature matrix},
    user1={$-$},
    sort={V_vec},
    type=symbols
}

\newglossaryentry{videoEmb}{
    name={video embedding},
    symbol={\mathcal{V}},
    description={video embedding tensor},
    user1={$-$},
    sort={V_cal},
    type=symbols
}

\newglossaryentry{fusedEmb}{
    name={fused embedding},
    symbol={\mathcal{Z}},
    description={cross-modal fused embedding tensor},
    user1={$-$},
    sort={Z_cal},
    type=symbols
}

\newglossaryentry{gateFunc}{
    name={gating function},
    symbol={\mathcal{H}},
    description={neural network gating function tensor},
    user1={$-$},
    sort={H_cal},
    type=symbols
}

\newglossaryentry{spkProfileList}{
    name={speaker profile list},
    symbol={\mathcal{E}},
    description={dynamic buffer of global speaker profiles},
    user1={$-$},
    sort={E_cal},
    type=symbols
}

\newglossaryentry{spkProfile}{
    name={speaker profile matrix},
    symbol={\vb{E}},
    description={matrix of registered global speaker profiles},
    user1={$-$},
    sort={E_vec},
    type=symbols
}

\newglossaryentry{spkEmb}{
    name={speaker embedding},
    symbol={\vb{e}},
    description={extracted speaker embedding vector},
    user1={$-$},
    sort={e_vec},
    type=symbols
}

\newglossaryentry{sigmoid}{
    name={sigmoid activation},
    symbol={\sigma},
    description={sigmoid non-linear activation function},
    user1={$-$},
    sort={z_sigma},
    type=symbols
}

\newglossaryentry{modelTrack}{
    name={tracking model},
    symbol={\mathcal{F}_{\text{track}}},
    description={speaker tracking system overall model},
    user1={$-$},
    sort={F_cal_track},
    type=symbols
}

\newglossaryentry{modelSS}{
    name={separation model},
    symbol={\mathcal{F}_{\text{SS}}},
    description={speaker separation neural network model},
    user1={$-$},
    sort={F_cal_ss},
    type=symbols
}

\newglossaryentry{modelVAD}{
    name={VAD model},
    symbol={\mathcal{F}_{\text{VAD}}},
    description={voice activity detection neural network model},
    user1={$-$},
    sort={F_cal_vad},
    type=symbols
}

\newglossaryentry{modelSPK}{
    name={speaker model},
    symbol={\mathcal{F}_{\text{SPK}}},
    description={speaker embedding extraction model},
    user1={$-$},
    sort={F_cal_spk},
    type=symbols
}

\newglossaryentry{nspkMax}{
    name={maximum concurrent speakers},
    symbol={K},
    description={Maximum number of overlapped speakers the separation model can represent},
    user1={$-$},
    sort={K_max},
    type=symbols
}

\newglossaryentry{seqLenOvlap}{
    name={overlap sequence length},
    symbol={L_{\text{ovl}}},
    description={Length of the overlapping region between consecutive segments},
    user1={$-$},
    sort={L_ovlap},
    type=symbols
}

\newglossaryentry{permElement}{
    name={permutation},
    symbol={\pi},
    description={A single permutation mapping from the permutation set},
    user1={$-$},
    sort={z_pi},
    type=symbols
}

\newglossaryentry{duration}{
    name={duration},
    symbol={T},
    description={duration of speech in seconds},
    user1={$-$},
    sort={T},
    type=symbols
}

\newglossaryentry{nWords}{
    name={number of words},
    symbol={W},
    description={total number of reference words},
    user1={$-$},
    sort={W},
    type=symbols
}

\newglossaryentry{muNormal}{
    name={Gaussian mean},
    symbol={{\mu}},
    description={Mean parameter of the truncated Gaussian distribution},
    user1={$-$},
    sort={z_mu},
    type=symbols
}

\newglossaryentry{sigmaNormal}{
    name={Gaussian standard deviation},
    symbol={{\sigma}},
    description={Standard deviation parameter of the truncated Gaussian distribution},
    user1={$-$},
    sort={z_sigma},
    type=symbols
}

\newglossaryentry{pairDist}{
    name={pairwise distance},
    symbol={d_{k', k}},
    description={pairwise distance between the trailing portion of speaker $k'$ in segment $\tau-1$ and the leading portion of speaker $k$ in segment $\tau$},
    user1={$-$},
    sort={d},
    type=symbols
}

\newglossaryentry{admissibleSet}{
    name={admissible pair set},
    symbol={\mathcal{G}},
    description={set of speaker pairs $(\ispk', \ispk) \in \activeSet_{\iseg-1} \times \activeSet_{\iseg}$ whose pairwise distance satisfies the stitching threshold $\stitchThreshold$},
    user1={$-$},
    sort={G_cal},
    type=symbols
}

\newglossaryentry{admissibleSetPrev}{
    name={admissible active set (previous)},
    symbol={\mathcal{A}^{*}_{\tau-1}},
    description={subset of $\activeSet_{\iseg-1}$ containing speakers that appear in at least one admissible pair in $\admissibleSet$},
    user1={$-$},
    sort={A_cal_star_prev},
    type=symbols
}

\newglossaryentry{admissibleSetCurr}{
    name={admissible active set (current)},
    symbol={\mathcal{A}^{*}_{\tau}},
    description={subset of $\activeSet_{\iseg}$ containing speakers that appear in at least one admissible pair in $\admissibleSet$},
    user1={$-$},
    sort={A_cal_star_curr},
    type=symbols
}

\newglossaryentry{permSetAdmissible}{
    name={admissible permutation set},
    symbol={\mathcal{P}^{*}},
    description={set of all one-to-one assignments $\perm \subseteq \admissibleSet$ between $\activeSet^{*}_{\iseg-1}$ and $\activeSet^{*}_{\iseg}$},
    user1={$-$},
    sort={P_cal_star},
    type=symbols
}

\newcommand{\nChn}{\glssymbol{nChn}}
\newcommand{\nspkMax}{\glssymbol{nspkMax}}
\newcommand{\seqLen}{\glssymbol{seqLen}}
\newcommand{\itime}{\glssymbol{itime}}

\newcommand{\ispk}{\glssymbol{ispk}}
\newcommand{\gspk}{\glssymbol{gspk}}
\newcommand{\Gspk}{\glssymbol{Gspk}}
\newcommand{\iseg}{\glssymbol{iseg}}

\newcommand{\seqLenOvlap}{\glssymbol{seqLenOvlap}}

\newcommand{\reals}{\glssymbol{reals}}

\newcommand{\permSet}{\glssymbol{permSet}}
\newcommand{\permSetWER}{\glssymbol{permSetWER}}
\newcommand{\activeSet}{\glssymbol{activeSet}}

\newcommand{\verificationThreshold}{\glssymbol{verificationThreshold}}
\newcommand{\stitchThreshold}{\glssymbol{stitchThreshold}}
\newcommand{\cossim}{\glssymbol{cossim}}
\newcommand{\distMetric}{\glssymbol{distMetric}}
\newcommand{\optPerm}{\glssymbol{optPerm}}
\newcommand{\optSpk}{\glssymbol{optSpk}}
\newcommand{\perm}{\glssymbol{perm}}

\newcommand{\sisdralpha}{\glssymbol{sisdralpha}}

\newcommand{\Loss}{\glssymbol{LossBase}}

\newcommand{\estimateScaler}{\glssymbol{estimateScaler}}

\newcommand{\mix}{\glssymbol{mix}}
\newcommand{\source}{\glssymbol{source}}

\newcommand{\estimate}{\glssymbol{estimate}}

\newcommand{\sourceSTFT}{\glssymbol{sourceSTFT}}

\newcommand{\estimateSTFT}{\glssymbol{estimateSTFT}}

\newcommand{\spkProfileList}{\glssymbol{spkProfileList}}
\newcommand{\spkProfile}{\glssymbol{spkProfile}}
\newcommand{\spkEmb}{\glssymbol{spkEmb}}

\newcommand{\modelTrack}{\glssymbol{modelTrack}}
\newcommand{\modelSS}{\glssymbol{modelSS}}
\newcommand{\modelVAD}{\glssymbol{modelVAD}}
\newcommand{\modelSPK}{\glssymbol{modelSPK}}

\newcommand{\duration}{\glssymbol{duration}}
\newcommand{\nWords}{\glssymbol{nWords}}

\newcommand{\substitution}[1][]{W_{#1}^S}
\newcommand{\deletion}[1][]{W_{#1}^D}
\newcommand{\insertion}[1][]{W_{#1}^I}
\newcommand{\leadingSeg}[1][]{\glssymbol{leadingSeg}_{#1}}
\newcommand{\trailingSeg}[1][]{\glssymbol{trailingSeg}_{#1}}

\newcommand{\admissibleSet}{\glssymbol{admissibleSet}}

%% file: Sections/0-abstract.tex
\begin{abstract}
Speaker tracking is the task of separating and following multiple speakers over time. It must address overlapped speech, speech onset and offset, talker identity, and time-varying speaker count. We propose a segment-online, modular framework for single- and multi-channel speaker tracking. The proposed system first performs speaker separation in each segment and computes speech activity via \gls{vad}. To generate speaker tracks over time, we introduce a two-stage sequential organization strategy: Overlap-based stitching for continuous grouping and memory-based speaker verification for discontinuous grouping. For speaker separation, we employ complex spectral mapping to estimate the real and imaginary spectrograms of underlying speakers. The proposed system achieves state-of-the-art segment-online tracking performance on the LibriCSS and AMI datasets. Our framework significantly reduces \gls{der} and \gls{cpwer} compared to other methods.
\end{abstract}


%% file: Sections/01-Introduction.tex
\section{Introduction}
\label{sec:introduction}

Speaker tracking is the broad task of separating overlapped talkers and following them over an extended period. Inspired by auditory scene analysis principles~\cite{bregman1994auditory}, we formulate speaker tracking as two related processes: simultaneous grouping or organization, and sequential organization (see also \cite{wang2006computational}). The simultaneous process deals with an acoustic mixture over a short time, or a segment of typically a few seconds long. When multiple speakers are active concurrently, simultaneous organization aims to separate this mixture into simultaneous streams, each corresponding to one speaker in the segment. On the other hand, sequential organization integrates simultaneous streams across segments to form speaker tracks, each corresponding to a distinct speaker over a longer period of time (such as a meeting). Sequential organization provides the mechanism of grouping the discontinuous utterances of the same talker.

Human listeners are remarkably adept at tracking speakers in noisy and dynamic environments~\cite{darwin2008listening}, which is essentially the cocktail party problem~\cite{cherry1957human}. In contrast, computational speaker tracking remains a formidable challenge~\cite{moraru2005experiments,watanabe20b_chime}. Real-world acoustic scenes include background noise and room reverberation, which degrade signal quality and obscure speech cues~\cite{wang2018supervised}. Overlapped speech from concurrent talkers further complicates the task by blending the talker characteristics of overlapped speakers. In addition, speaker tracking has to maintain talker identity across discontinuous time periods, hence requiring a form of memory for extended durations and speaker modeling, and detecting new speakers.

Speaker tracking is related to, but broader than, speaker separation and speaker diarization. Speaker separation focuses on isolating concurrent voices within a specific segment, corresponding to simultaneous organization~\cite{wang2018supervised}. Speaker diarization is capable of linking same-talker utterances across time~\cite{anguera2012speaker,park2022review}. A recent  diarization study also uses the term speaker tracking to denote following many speakers over time~\cite{kwon26_interspeech}. However, diarization systems usually do not deal with overlapped speech, or label it as belonging to more than one talker without separating the underlying signals.  Speaker tracking inherently addresses both tasks, separating overlapped speech and diarizing separated speech. The objective of speaker tracking is to generate the separated and enhanced speech signal of each speaker throughout their presence in an acoustic scene.

Deep-learning-based speaker separation has witnessed major advances in recent years~\cite{wang2018supervised}. Models such as deep clustering~\cite{hershey2016deep} and \gls{pit}~\cite{kolbaek2017multitalker} have successfully addressed the permutation ambiguity issue in talker-independent separation in monaural settings.  Subsequently, time-domain methods, including Conv-TasNet~\cite{luo2019conv}, were introduced to replace the short-time Fourier transform (STFT) with learnable encoders. A recent trend is the resurgence of \gls{tf} domain approaches based on \gls{csm}, which trains a \gls{dnn} to predict real and imaginary spectrograms~\cite{williamson2016complex}. \Gls{csm} can fundamentally represent both signal magnitude and phase, and hence naturally encodes inter-channel phase relations of sound sources in multi-channel recordings, providing an important spatial dimension. \Gls{csm} models, such as TF-GridNet~\cite{wang2022tfjournal}, SpatialNet~\cite{quan2023spatialnet}, and TF-CrossNet~\cite{kalkhorani2024crossnet}, have achieved state-of-the-art speaker separation performance.

To move toward real-world applicability, subsequent work has proposed \gls{css} to separate speaker streams in a conversational setting~\cite{chen2020continuous, wang2021multi,han2021continuous,wang2022localization,shin2025tf,kim25h_interspeech}. \Gls{css} processes a sliding sequence of partially overlapped segments and outputs a fixed number of overlap-free streams, irrespective of the total speaker count. \Gls{css} addresses simultaneous grouping within each segment and stitches the same-talker signal across consecutive segments, but does not perform sequential grouping across discontinuous segments. Consequently, a single output stream may contain speech from multiple speakers at different times. Such sequential grouping requires a separate diarization backend. Previous systems have combined \gls{css} and speaker diarization~\cite{yoshioka2019advances,rajSLTIntegration,von2024meeting, abramovski2025summary}. Yoshioka et al.~\cite{yoshioka2019advances} integrate \gls{css} with online audio-visual diarization in a separate-recognize-diarize pipeline, but rely on a camera feed and a prior list of meeting attendees. Von Neumann et al.~\cite{von2024meeting} combine \gls{css} with offline d-vector clustering and ASR-supported speaker turn detection, which achieves strong \gls{cpwer} on LibriCSS, but operates offline. 

Traditional speaker diarization pipelines use \gls{vad}, embedding vector extraction, and clustering~\cite{Sell2018dihard, Diez2018, Snyder2019}. While effective offline, these systems often assume a fixed speaker count and struggle with speaker overlaps. Overlap-aware methods, including \gls{eend} models, jointly perform detection and attribution~\cite{Fujita2019Interspeech, fujita2019end, horiguchi2022encoder}, with extensions to unknown speaker counts~\cite{horiguchi20_interspeech, han2021bw, liang2025ls}. Nevertheless, diarization models do not separate overlapped speech signals, limiting their utility for downstream tasks such as ASR.

An alternative to \gls{css}, termed \gls{ssnd}~\cite{taherian2024multi}, employs an offline \gls{eend} to estimate speech activities and condition a downstream speaker separation model. Assuming a maximum of two overlapped talkers at each time-frame, the speaker separation model in this framework outputs two continuous streams regardless of the total speaker count. \Gls{ssnd} extracts speaker embeddings from non-overlapped segments identified by the diarization module. These embeddings, combined with the diarization results, guide the assignment of separated signals to the output streams as a post-processing step. DCF-DS~\cite{niu2025dcf} further integrates diarization and separation through a deep cascade fusion mechanism, allowing each module to condition on the other, and was recently extended to exploit multi-channel spatial information~\cite{niu26b_interspeech}. However, like \gls{ssnd}, DCF-DS is an offline model and needs prior speaker diarization results. 

Despite these advances, a significant research gap remains between the current separation and diarization results and online speaker tracking. Existing systems often rely on offline backends for sequential organization or require prior speaker enrollment to guide separation.  Little prior research aims to  provide a unified system for single-channel and multi-channel segment-online speaker tracking without assuming a fixed number of speakers. 
Addressing these limitations would require a broad framework that deals with both simultaneous and sequential organization. Sequential organization would need to be mediated by a form of working memory, referring to temporary storage of information for performing complex tasks~\cite{baddeley2025memory}.

This paper tackles the speaker tracking challenge by developing a segment-online, modular framework capable of both simultaneous and sequential organization with an unknown number of speakers. Simultaneous organization within a segment will be conducted through speaker separation based on complex spectral mapping, producing simultaneous streams. We propose to address sequential organization across different segments in two stages. The first stage performs overlap-based stitching to organize simultaneous streams, to extend simultaneous streams across consecutive segments. The second stage sequentially organizes discontinuous simultaneous streams into speaker tracks by introducing a working memory mechanism based on speaker verification. The proposed two-stage approach addresses the limitations of cascade systems: Stitching based on spectral overlap maintains the temporal continuity of speech from one segment to the next, and memory-driven verification maintains talker identity over extended periods of no speech activity. Unlike the SSND and DCF-DS models, our framework groups the discontinuous speech signals of the same talker without a separate offline backend or prior speaker enrollment. We present both single-channel and multi-channel tracking algorithms. In single-channel scenarios, tracking relies on spectral, temporal, and talker cues. In the multi-channel setup, tracking further exploits spatial cues to improve both simultaneous and sequential organization. We systematically evaluate the proposed framework on the LibriCSS~\cite{chen2020continuous} and AMI~\cite{ami} datasets.

The primary contributions of this paper are summarized as follows. First, we propose a segment-online speaker tracking framework that integrates speaker separation and sequential organization in a modular fashion. Second, we propose a two-stage sequential organization approach, where a working memory mechanism is utilized to organize temporally-discontinuous utterances into distinct speakers. Third, the proposed single-channel and multi-channel speaker tracking algorithms advance the state-of-the-art performance on the LibriCSS and AMI datasets.

The rest of the paper is organized as follows. Section~\ref{sec:proposed_method} describes the proposed speaker tracking framework. Section~\ref{sec:experimental_setup} presents the experimental setup. Section~\ref{sec:results} describes the evaluation and comparison results. Section~\ref{sec:conclusion} concludes the paper.

%% file: Sections/02-SystemOverview.tex

%% file: Sections/03-Methodology.tex
\begin{figure*}[th]
    \centering
    \includegraphics[width=0.95\linewidth]{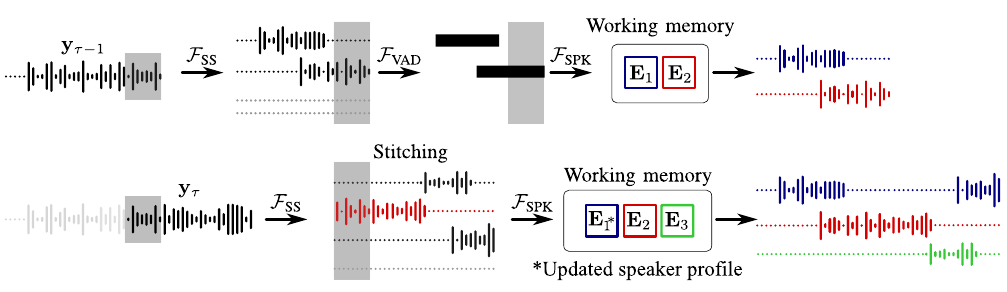}
    \caption{Illustration of the proposed modular speaker tracking framework. The architecture processes the input as sliding segments in different stages: speaker separation for simultaneous organization within each segment, VAD to compute activity traces, overlap-based stitching to group simultaneous streams across consecutive segments, and a working memory mechanism based on speaker embeddings to sequentially organize discontinuous streams into speaker tracks. Black waveforms denote the mixture and separated speakers within individual segments before sequential organization. A color denotes a distinct speaker track after sequential organization. Gray regions mark overlap intervals.}
    \label{fig:overall}
\end{figure*} 

\section{Model description}
\label{sec:proposed_method}

A varying number of active speakers poses a considerable challenge to speech separation, as \gls{dnn} models would require a fixed output dimension. In real-world environments, speakers may enter or exit an auditory scene at arbitrary times, resulting in varying numbers of concurrent sources and disjoint speaker tracks. A robust tracking system must therefore function without prior knowledge of the total number of speakers in the scene, and maintain consistent speaker identities across the entire session.

To achieve this, the proposed tracking system, denoted by $\modelTrack$, maps the input observations $\mix({\itime}) \in \reals^{\nChn}$ in a session to a set of estimated speaker streams $\{\estimateScaler_{\gspk}(\itime)\}_{\gspk=0}^{\Gspk(\itime)}$, where $\nChn$ is the number of microphones, $\itime$ is the discrete time index, $\gspk$ is the globally consistent speaker index, and $\Gspk(\itime)$ is the total number of distinct speakers tracked up to time $\itime$

\begin{equation}
\label{eq:tracking_formulation}
     \left\{ \estimateScaler_{\gspk}(\itime) \right\}_{\gspk=0}^{\Gspk(\itime)} = \modelTrack \big( \mix({\itime}) \big).
\end{equation} 
In this formulation, $\Gspk(\itime)=0$ indicates that no active speaker has been tracked by time $\itime$, e.g., during the session beginning, and $\estimateScaler_{0}(\itime)$ denotes the nonspeech background signal.

When \(\nChn = 1\) (single-channel), the system must rely on spectro-temporal patterns to distinguish speakers. In the multi-channel case, \(\nChn > 1\), the system additionally exploits spatial cues, such as inter-channel phase difference, to spatially filter and track sources. The system must maintain global speaker correspondence to ensure that the estimate $\estimateScaler_{\gspk}(\itime)$ assigned to speaker $\gspk$ corresponds to the same talker at all times, regardless of overlaps, interruptions, or silences. Sequential organization is required to prevent talker label switching across segments, particularly when multiple speakers are active (i.e., talking) simultaneously.

\subsection{Speaker separation for simultaneous grouping}

We propose a modular segment-online architecture starting with speaker separation, as shown in \autoref{fig:overall}. The input audio is divided into overlapped segments indexed by $\iseg$. A separation model $\modelSS$ estimates a fixed number of candidate streams
\begin{equation}
    \left\{ \estimate_{\iseg,\ispk} \right\}_{\ispk=1}^{\nspkMax} = \modelSS \big( \mix_{\iseg} \big),
\end{equation} 
where $\ispk$ indexes speakers within segment $\iseg$, $\nspkMax$ is a hyperparameter denoting the maximum number of separated speakers the speaker separation model can represent in a segment, $\mix_{\iseg} \in \reals^{\nChn \times \seqLen}$ is the mixture in segment $\iseg$, $\seqLen$ denotes the number of samples in a segment, and $\estimate_{\iseg,\ispk} \in \reals^{\seqLen}$ denotes the estimated signal for speaker $\ispk$ in segment $\iseg$.  When a segment contains fewer than $\nspkMax$ talkers, some of the candidate streams will be a zero signal.

While the speaker separation model can effectively separate simultaneous talkers within each segment, it does not maintain speaker identity across segments. Therefore, additional processing is required to associate separated streams over time to ensure consistent speaker labeling. This association relies on both continuous and discontinuous organization, as detailed in Sec.~\ref{sec:sequential_grouping} and Sec.~\ref{subsec:working_memory}.

To resolve permutation ambiguity during the training, we utilize \gls{pit}~\cite{kolbaek2017multitalker}. Since some speakers may be silent, the training objective should take into account inactive candidate streams. We therefore propose \gls{si-sa-sdr} as the primary loss function. This formulation extends \gls{sa-sdr}~\cite{von2022sa} by introducing scale-invariance as follows:

\begin{equation}
\label{eq:si-sa-sdr}
\loss{SI-SA-SDR} = -10\log_{10} \frac{\sum_{\ispk=1}^{\nspkMax} \norm{\sisdralpha \source_{\iseg,\ispk}}_2^2}{\sum_{\ispk=1}^{\nspkMax} \norm{\sisdralpha \source_{\iseg,\ispk} - \estimate_{\iseg,\ispk}}_2^2},
\end{equation}
where $\source_{\iseg,\ispk}$ is the clean (reference) signal for speaker $\ispk$ in segment $\iseg$, $\norm{\cdot}_2$ is the $L_2$ norm operator, and $\sisdralpha$ is a single scalar value shared across all $\nspkMax$ speakers, calculated as

\begin{equation}
\sisdralpha = \frac{\sum_{\ispk=1}^{\nspkMax} \estimate_{\iseg,\ispk}\tran \source_{\iseg,\ispk}}{\sum_{\ispk=1}^{\nspkMax} \source_{\iseg,\ispk}\tran \source_{\iseg,\ispk}}.
\end{equation}

To further suppress residual energy in inactive outputs, we include a magnitude reconstruction loss. To ensure consistency between the complex domain predictions and the resulting waveforms, we compute the magnitude loss of the reconstructed signals:
\begin{equation}
\label{eq:mag_loss}
\loss{Mag} = \sum_{\ispk=1}^{\nspkMax} \norm{\, \abs{\sourceSTFT_{\iseg,\ispk}} - \abs{\estimateSTFT_{\iseg,\ispk}} \,}_1,
\end{equation} 
where $\abs{\sourceSTFT_{\iseg,\ispk}}$ and $\abs{\estimateSTFT_{\iseg,\ispk}}$ denote the magnitude of reference and estimated spectrograms in segment $\iseg$ of speaker $\ispk$, respectively, and $\norm{\cdot}_1$ is the $L_1$ norm operator.

The final loss is the sum of the two terms:

\begin{equation}
\Loss = \loss{SI-SA-SDR} + \loss{Mag}.
\end{equation}

This combined objective yields stable training and strong separation quality and downstream \gls{asr} performance under reverberant, multi-talker conditions.

\subsection{Voice activity detection}

Following separation, we compute frame-level activity traces for each separated source using a pre-trained \gls{vad} model, denoted as $\modelVAD$ in ~\autoref{fig:overall}. These activity labels are used to remove speaker separation leakage errors, determine which speakers are active in the overlap interval between consecutive segments, and to restrict comparisons during overlap-based stitching. For the final diarization output, the \gls{vad} traces of each speaker from all segments are concatenated.

\subsection{Sequential organization in consecutive segments via stitching}
\label{sec:sequential_grouping}

The separation stage yields a set of local, segment-level speaker estimates. To form coherent speaker tracks, these local estimates must be sequentially organized. We propose a two-stage sequential organization strategy that addresses both short- and long-term sequential organization.

For consecutive segments $\mix_{\iseg-1}$ and $\mix_{\iseg}$, we utilize the overlapped interval of length $\seqLenOvlap$ samples to enforce temporal continuity. Since the acoustic realization of a separated speaker should remain stable over this short interval, we perform sequential grouping by matching the separated waveforms directly; in other words, the overlapped interval provides the ``thread'' for a stitching process.

Specifically, let $\estimate_{\iseg, \ispk} \in \reals^{\seqLen}$ denote the estimated signal for speaker $\ispk \in \{1, \ldots, \nspkMax\}$ in segment $\iseg$. We denote the first $\seqLenOvlap$ samples of $\estimate_{\iseg, \ispk}$ as $\leadingSeg[\iseg, \ispk] \in \reals^{\seqLenOvlap}$ and the last $\seqLenOvlap$ samples as $\trailingSeg[\iseg, \ispk] \in \reals^{\seqLenOvlap}$. 
Let $\activeSet_{\iseg-1}$, $\activeSet_{\iseg} \subseteq \{1, \ldots, \nspkMax\}$ denote the sets of active speakers separated within the overlapped interval by $\modelSS$ in segments $\iseg-1$ and $\iseg$, respectively. For each speaker pair $(\ispk', \ispk) \in \activeSet_{\iseg-1} \times \activeSet_{\iseg}$, we first compute the pairwise distance
\begin{equation}
    \label{eq:stitching}
    d_{\ispk', \ispk} = \distMetric\big(\trailingSeg[\iseg-1, \ispk'],\, \leadingSeg[\iseg, \ispk]\big),
\end{equation}
where $\distMetric$ is a distance metric. 
If $d_{\ispk', \ispk} > \stitchThreshold$, where $\stitchThreshold$ denotes the threshold of acceptable dissimilarity, the pair $(\ispk', \ispk)$ is removed. Afterwards, we remove speaker $\ispk$ from $\activeSet_{\iseg}$ if all the pairs involving $\ispk$ are removed, and speaker $\ispk'$ likewise. 

Next, we construct a set of all possible stitches $\permSet$ over the remaining active speakers in $\activeSet_{\iseg}$ and $\activeSet_{\iseg-1}$, where each $\perm \in \permSet$ is a stitch forming a specific pairing or permutation between all the remaining speakers in $\activeSet_{\iseg}$ and the remaining ones in $\activeSet_{\iseg-1}$. The optimal stitch $\optPerm$ is selected as
\begin{equation}
    \optPerm = \mathop{\argmin}_{\perm \in \permSet} \sum_{(\ispk', \ispk) \in \perm} d_{\ispk', \ispk}.
\end{equation}

For each pair $(\ispk', \ispk) \in \optPerm$, the separated stream of speaker $\ispk$ in segment $\iseg$ is stitched to that of speaker $\ispk'$ in segment $\iseg-1$. 
Specifically, $\trailingSeg[\iseg-1,\ispk']$ and $\leadingSeg[\iseg,\ispk]$ are combined via  an overlap-and-add technique. To smooth segment boundaries, we first apply complementary linear ramps that fade out $\trailingSeg[\iseg-1,\ispk']$ and fade in $\leadingSeg[\iseg,\ispk]$, and then sum the two ramped signals to form the new speaker signal in the overlapped interval. The remaining non-overlapped portion of $\estimate_{\iseg,\ispk}$ is appended to produce the stitched stream.
All active streams in segment $\iseg$ that are not stitched to those in $\iseg - 1$ will be further organized as described in Sec.~\ref{subsec:working_memory}.

Although the idea of stitching is intuitive and has been used previously~\cite{yoshioka18_interspeech,chen2020continuous}, we cannot find a quantitative description and there are various ways of implementing the stitching idea.
The stitching process described above is accurate and computationally efficient for continuous grouping. But it fails when speakers pause longer than $\seqLenOvlap$ or are not active in the overlapped interval, necessitating discontinuous grouping.

\subsection{Sequential organization in disjoint segments via working memory}
\label{subsec:working_memory}

In human cognition~\cite{baddeley2025memory} \gls{wm} serves to temporarily hold and manipulate information necessary for complex reasoning and tracking. Motivated by this form of human memory, our framework incorporates a \gls{wm} mechanism to perform discontinuous organization. The primary role of this mechanism is to maintain speaker identity beyond consecutive segments. While stitching effectively handles sequential grouping in consecutive segments, it fails when a speaker pauses for longer than a segment or is inactive in overlapped intervals.
To perform sequential grouping in disjoint segments, the \gls{wm} mechanism includes dynamically formed speaker models $\modelSPK$. When a speaker stream in the current segment $\iseg$ cannot be grouped to earlier speakers via stitching, we extract a speaker embedding vector
\begin{equation}
    \spkEmb_{\iseg,\ispk} = \modelSPK \big( \estimate_{\iseg,\ispk} \big). 
\end{equation}
The model $\modelSPK$ is a pre-trained speaker encoder (e.g., ReDimNet~\cite{yakovlev2024redimenet}) that maps variable-length audio segments to a fixed-dimensional discriminative embedding space.

The proposed working memory mechanism maintains a set of global speaker profiles $\spkProfileList(\itime) = \{\spkProfile_{\gspk}\}_{\gspk=1}^{\Gspk(\itime)}$. The newly extracted embedding $\spkEmb_{\iseg,\ispk}$ is compared against all the profiles in the \gls{wm} using the cosine similarity score:
\begin{equation}
\label{eq:cosine_sim}
    \cossim \big( \spkEmb_{\iseg,\ispk}, \spkProfile_{\gspk} \big) = \frac{ \spkEmb_{\iseg,\ispk}\tran \spkProfile_{\gspk} }{ \lVert \spkEmb_{\iseg,\ispk} \rVert_{2} \, \lVert \spkProfile_{\gspk} \rVert_{2} }.
\end{equation}

For each unstitched speaker in segment $\iseg$, we identify the most similar global speaker profile as
\begin{equation}
    \optSpk = \mathop{\argmax}_{\gspk \in \{1, \ldots, \Gspk(\itime)\}}
    \cossim\big(\spkEmb_{\iseg,\ispk},\spkProfile_{\gspk}\big).
\end{equation}
If $\cossim\big(\spkEmb_{\iseg,\ispk},\spkProfile_{\optSpk}\big) \geq \verificationThreshold$,
the stream is assigned to speaker $\optSpk$ and the profile $\spkProfile_{\optSpk}$ is updated using the moving average method (denoted as $\spkProfile_{1*}$ in \autoref{fig:overall}); otherwise, a new speaker profile is initialized and added to the \gls{wm} (e.g., $\spkProfile_3$ in \autoref{fig:overall}). By dynamically updating these profiles, the \gls{wm} acts as an adaptable speaker buffer. This \gls{wm} mechanism enables the system to maintain speaker identities across interruptions and silences without prior knowledge of the total number of speakers, bridging the gap between simultaneously organized streams and coherent speaker tracks.

Existing buffer-based methods commonly store mixture signals or embeddings for repeated evaluations. Examples include the speaker-tracing buffer in SA-EEND~\cite{xue2021online}, block-wise \gls{eend}~\cite{horiguchi2022online}, and the arrival-order speaker cache in streaming Sortformer~\cite{medennikov25_interspeech}. These methods concatenate incoming features with buffered representations and reprocess the entire sequence for each new chunk. This repeated processing increases computational cost and introduces variable-length inputs that require variable chunk-size training~\cite{horiguchi2022online}. In addition, feature buffers may contain overlapped speech depending on the frame selection policy, and the total number of output speakers may be restricted. 
In contrast, our \gls{wm} mechanism maintains compact, speaker-specific embedding profiles extracted from separated speech. Our framework only limits the maximum number of simultaneously active speakers (i.e., $\nspkMax$), and allows the total number of global speaker tracks to grow dynamically. Leveraging continuous grouping via stitching and discontinuous grouping via speaker profiling, our approach completely avoids repeated inference over past segments.

\subsection{Implementation details}

 We adopt TF-CrossNet \cite{kalkhorani2024crossnet} as the speaker separation model. TF-CrossNet performs \gls{csm} in the \gls{tf} domain and predicts the real and imaginary parts of each speaker's spectrogram. The TF-CrossNet architecture comprises five modules: (i) an encoder layer to project mixture \gls{ri} representations; (ii) a \gls{gmhsa} module to model cross‑frame and cross‑band dependencies; (iii) a cross‑band module to capture inter‑frequency correlations across all channels; (iv) a narrow‑band module to model fine-grained structure within adjacent frequency bins; and (v) a decoder that yields separated spectrograms, which are converted back to time-domain waveforms via inverse STFT \cite{kalkhorani2024crossnet}. TF-CrossNet also incorporates \gls{rcpe} to mitigate performance degradation on long utterances and improve temporal generalization \cite{kalkhorani2024crossnet}; however, we do not utilize \gls{rcpe} here as we operate on relatively short segments.  See \cite{kalkhorani2024crossnet} for further details. For evaluations on the AMI dataset, the speaker separation model is further fine-tuned on the AMI training set for 10 epochs following training on a simulated dataset (see \autoref{sec:datasets}).

We utilize three pre-trained VAD models: Pyannote~\cite{bredin2020pyannote}, Silero VAD\footnote{\url{https://github.com/snakers4/silero-vad}}, and WebRTC\footnote{\url{https://github.com/wiseman/py-webrtcvad/}}. Pyannote employs a neural sequence labeling architecture trained on a diverse set of conversational recordings and provides frame-level speech activity scores. Silero is a lightweight, real-time model optimized for low-latency applications, producing binary decisions at high temporal resolution. WebRTC is a signal processing based model that classifies short frames as speech or non-speech using a Gaussian mixture model, and is computationally efficient.

For the stitching mechanism  for continuous grouping, we use the negative \gls{si-sdr} as the distance measure $\distMetric$ in \autoref{eq:stitching}, and set the threshold $\stitchThreshold$ to $-\SI{5}{dB}$.

We experiment with five speaker verification models:
 NeMo~\cite{kuchaiev2019nemo},
 ReDimNet~\cite{yakovlev2024redimenet},
 ReDimNet2~\cite{yakovlev2026redimnet2},
 SpeechBrain~\cite{ravanelli2021speechbrain}, and
 Wespeaker~\cite{wang2023wespeaker}. 
 Each model extracts a fixed-dimensional embedding from a speech segment, which is then compared to previously-tracked speakers the using cosine similarity in \autoref{eq:cosine_sim}.  If the maximal similarity score exceeds a predefined threshold, the speech segment is added to the track of the closest talker; otherwise, a new track is initiated. 
To determine the optimal threshold $\verificationThreshold$ for each model, we evaluate speaker verification accuracy on 3000 speaker pairs randomly sampled from LibriSpeech development subset~\cite{panayotov2015librispeech}, with positive (same-speaker) and negative (different-speaker) pairs sampled with equal probability. We then perform a grid search over $\verificationThreshold$ ranging from 0.05 to 0.95 in steps of 0.05, and select the threshold that yields the highest verification accuracy.
For ReDimNet, we utilize the pre-trained M-sized model, which is trained on the combined VoxBlink2~\cite{lin24j_interspeech}, VoxCeleb2~\cite{chung2018voxceleb2}, and CN-Celeb~\cite{fan2020cnceleb} datasets\footnote{~\url{https://github.com/IDRnD/ReDimNet/}}, and for ReDimNet2 we select the pre-trained B6 model which is trained on VoxCeleb2 dataset\footnote{~\url{https://github.com/PalabraAI/redimnet2/}}. In all cases, the speaker model weights are frozen and we do not fine-tune them.

%% file: Sections/04-ExperimentalSetup.tex
\section{Experimental setup}
\label{sec:experimental_setup}

\subsection{Datasets}
\label{sec:datasets}

To train the proposed speaker tracking system, we generate simulated mixtures that reflect the acoustic characteristics of two evaluation datasets: the LibriCSS~\cite{chen2020continuous} and the AMI~\cite{ami} meeting datasets. For both configurations, each mixture includes at most four speakers drawn from the \textit{train-clean-100} and \textit{train-clean-360} subsets of LibriSpeech~\cite{panayotov2015librispeech}. \Glspl{rir} are synthesized using the gpuRIR~\cite{diaz2021gpurir} toolkit, with 100,000 and 2,500 rooms simulated for the training and validation sets, respectively.

For the LibriCSS simulation, room dimensions are sampled uniformly with lengths and widths ranging from 3 to 8~\si{m} and heights from 3 to 4~\si{m}. The \gls{rt60} is drawn from a truncated Gaussian distribution ($\mu=0.4$~\si{s}, $\sigma=0.12$~\si{s}) bounded between 0.1 and 0.6~\si{s}. We construct a 7-channel microphone array corresponding to the LibriCSS hardware geometry (radius 4.25~\si{cm}), positioned within 1.0~\si{m} of the room center at an elevation between 1.0 and 1.5~\si{m}. Speakers are distributed at radial distances of 0.5 to 4.5~\si{m} from the array, with heights ranging from 1.0 to 1.5~\si{m}. To ensure sufficient spatial separability, a minimum azimuth difference of $5^\circ$ is maintained across all sources. 

Following the LibriCSS recipe \cite{chen2020continuous}\footnote{Available at:~\url{https://github.com/jsalt2020-asrdiar/jsalt2020_simulate}}, we simulate meeting-style conversations where two to four speakers each provide two utterances from the LibriSpeech training set \cite{panayotov2015librispeech}. 
Each conversation is simulated as follows. First, an utterance is randomly drawn from the set of the utterances from all speakers in a conversation. Then a silence of a random duration between 0.5~s to 3.0~s is added with a 26\% probability. If a silence is added, the next utterance is drawn from the set of the remaining utterances. If no silence is added, the next utterance is drawn from the set of the remaining utterances with the condition that the drawn utterance cannot be from the same talker as the current utterance. Then the current and next utterance are mixed with a speaker overlap ratio randomly selected from 0\% to 45\%. This process continues until all the utterances of the set are used up. The simulated meetings are partitioned into \SI{8}{\second} segments for speaker separation model training.

We evaluate speaker tracking in both session- and section-wise scenarios. 
In session-wise evaluation, each ten-minute mini-session signal is processed 
directly. For section-wise evaluation, we utilize the pre-segmented LibriCSS sections. 
These sections range from 5 to 120 seconds in length and contain between 2 
and 10 utterances from up to 8 distinct speakers.

For AMI data simulation, we generate \glspl{rir} modeling standard conference rooms. Room dimensions are sampled uniformly with lengths and widths spanning 3 to 10~\si{m} and heights from 3 to 4~\si{m}. \Gls{rt60} is drawn from the same truncated Gaussian distribution but bounded between 0.2 and 0.7~\si{s} to reflect heavier reverberation. We simulate a single microphone positioned near the room center at a vertical height of 0.7 to 0.85~\si{m} to represent a table-mounted microphone. Up to four simultaneous speakers are distributed around the microphone with radii from 0.75 to 2.5~\si{m} and elevations from 1.1 to 1.3~\si{m}, representing seated participants. The $5^\circ$ minimum azimuth separation is also enforced across all localized sources. 

In both simulation protocols, diffuse background noise is generated using the method described in Habets \etal~\cite{habets2008generating}, with single-channel ambient noises drawn from the REVERB challenge dataset \cite{kinoshita2013reverb}. The mixture \gls{snr} is randomly selected between 5 and 20~\si{dB}. To support both single- and multi-channel configurations, we adapt array usage accordingly: for multi-channel models, the full microphone array is utilized, whereas for single-channel models, we randomly select a single channel for training and evaluate using the primary reference channel.

\subsection{Evaluation metrics}
\label{subsec:metrics}
We use \gls{der} to assess diarization performance, and speaker-attributed \gls{cpwer} to assess ASR performance.
\Gls{der} quantifies diarization accuracy as the fraction of reference speech duration attributed to \gls{mi}, \gls{fa}, or \gls{cf} errors:
\begin{equation}
\label{eq:der}
\gls{der} = \frac{\duration_{\gls{mi}} + \duration_{\gls{fa}} + \duration_{\gls{cf}}}{\duration_{\text{total}}},
\end{equation}
where $\duration_{\gls{mi}}$, $\duration_{\gls{fa}}$, and $\duration_{\gls{cf}}$ are the durations of missed speech, false alarm, and speaker confusion, respectively, and $\duration_{\text{total}}$ is the total reference speech duration. A 0-\si{s} collar tolerance is applied in all evaluations.

\Gls{wer}~\cite{chen2020continuous} measures transcription accuracy without enforcing speaker identity. Reference and hypothesis words are pooled across all speakers, and \gls{wer} is defined as
\begin{equation}
\label{eq:wer}
\gls{wer} = \frac{\substitution + \deletion + \insertion}{\nWords},
\end{equation}
where $\substitution$, $\deletion$, and $\insertion$ are the numbers of substitution, deletion, and insertion word errors, respectively, and $\nWords$ is the total number of reference words.

Metric \gls{cpwer}~\cite{watanabe20b_chime} extends \gls{wer} to the multi-talker setting by 
jointly optimizing all speaker assignments. The utterances of each talker are first concatenated 
into a single stream for both reference and hypothesis. \Gls{wer} is then evaluated over 
all possible speaker assignments $\perm \in \permSetWER$ of the hypothesis, and the minimum 
is taken as the final score:

\begin{equation}
\label{eq:cpwer}
    \gls{cpwer} = \min_{\perm \in \permSetWER} \frac{\sum_{(\gspk, \gspk') \in \perm} 
    \left( \substitution[\gspk, \gspk'] + \deletion[\gspk, \gspk'] + 
    \insertion[\gspk, \gspk'] \right)}{\sum_{\gspk} W_{\gspk}},
\end{equation} %
where $\permSetWER$ is the set of all pairings between reference and hypothesis 
speakers, $(\gspk, \gspk') \in \perm$ denotes a hypothesized pair of reference speaker $\gspk$ 
and hypothesis speaker $\gspk'$, $\substitution[\gspk, \gspk']$, $\deletion[\gspk, \gspk']$, 
and $\insertion[\gspk, \gspk']$ are the substitution, deletion, and insertion errors 
when reference speaker $\gspk$ is matched to hypothesis speaker $\gspk'$, and 
$W_{\gspk}$ is the number of words spoken by reference speaker $\gspk$.

To properly penalize speaker count mismatches during evaluation, we pad the smaller speaker set with empty speaker streams prior to computing metrics. Specifically, we add zero-valued signals for \gls{der} and empty transcripts for \gls{cpwer}. This ensures unmatched hypothesis speakers count as false alarms or insertions, and unmatched reference speakers count as missed speech or deletions.

%% file: Sections/05-EvaluationResults.tex
\section{Evaluation results and comparisons}
\label{sec:results}

\subsection{Results on LibriCSS}
We first evaluate our system on the LibriCSS dataset using \gls{der}, and \gls{cpwer} as metrics. \autoref{tab:diarization_results} reports \gls{der} results across all overlap conditions. Existing comparison approaches fall into two categories: (i) systems that apply a separation or beamforming front-end followed by diarization, and (ii) systems that directly perform diarization, i.e., without front-end separation.

\input{Tables/diarization}

The first category includes mask-based MVDR~\cite{chen2020continuous}, MIMO-BF-MISO~\cite{taherianICASSP24}, and MISO-BF-MISO~\cite{wang2021multi}, which combine spatial filtering and clustering-based diarization (i.e., x-vector + spectral clustering). While effective in reducing overlap-caused errors, these systems are not designed for online operation. In addition, LlaSep~\cite{lanzendoerfer26b_interspeech} performs generative separation using an autoregressive audio language model and determines speaker activity from the separated streams via energy-based thresholding, achieving an average \gls{der} of \SIpct{23.43}.
The second category consists of diarization models without speaker separation, including x-vector + SC, AHC, VBx~\cite{rajSLTIntegration}, and SC-EEND~\cite{taherian2024multi}. These models are simpler but tend to degrade more severely under high overlap conditions due to the absence of a separation stage.
Competitive methods include RPN and TS-VAD~\cite{rajSLTIntegration}, which leverage speaker-specific enrollment and target speaker detection networks. These methods achieve low \gls{der} (e.g., \SI{7.6}{\percent} for TS-VAD) but rely on oracle knowledge of the number of speakers, an assumption not made in our framework.

The proposed single-channel model achieves 4.63\% \gls{der}. This represents an absolute reduction of more than 13.5\% relative to the x-vector + SC baseline~\cite{rajSLTIntegration} (\SI{18.19}{\percent}). The proposed seven-channel model further reduces the \gls{der} to the new state-of-the-art score of {3.38\%}, outperforming the previous best MC-EEND~\cite{taherian2024multi} (\SI{4.68}{\percent}), despite MC-EEND operating offline. 

To isolate the effects of speaker separation from subsequent sequential organization, we report results under an ``Oracle'' condition. In this setup, we utilize reference dry signals (\textit{each\_spk} files) from LibriCSS as separated speech for two-stage sequential grouping. The oracle results provide the lower bound on errors caused by \gls{vad}, speaker models, and the ASR backend. As shown in \autoref{tab:diarization_results}, our 7-channel \gls{der} of 3.38\% is very close to the oracle \gls{der} of 2.98\%, demonstrating the high separation quality of TF-CrossNet.

\input{Tables/LibriCSS_WER.tex}

For the ASR task, we report speaker-attributed \gls{cpwer} in \autoref{tab:LibriCSS_cpWER}. While standard LibriCSS evaluations often prioritize \gls{wer}, \gls{cpwer} provides a more meaningful assessment of speaker tracking by penalizing sequential grouping errors. Our proposed single-channel method achieves a \gls{cpwer} of {9.25\%} with E2E \gls{asr} and {7.20\%} with E2E-SSL \gls{asr}. Our seven-channel method further reduces \gls{cpwer} to {5.32\%} and {4.39\%} respectively, surpassing the MIMO-BF-MISO baseline (\SI{8.76}{\percent}) and achieving competitive performance to the offline SSND model (5.13\% and 3.22\%).

Compared to DCF-DS~\cite{niu2025dcf}, our segment-online system avoids full-utterance context dependencies. DCF-DS requires a {global diarization prior} to segment the audio before window-based processing. Furthermore, its stitching relies on global speaker indices to reconstruct continuous streams from local separation results. Compared to TMS~\cite{liu2026tms}, our system generates multiple tracks simultaneously without requiring semantic cues. TMS runs an online ASR system on the raw mixture to generate preliminary transcripts as text prompts. It then isolates the acoustic signal that aligns with the prompt, treating other voices as background interference.  This method involves a large language model and operates offline.

\subsection{Results on AMI}
\label{sec:results_ami}
We now evaluate our system on the AMI IHM-Mix meeting corpus using \gls{der} as the metric. Compared to LibriCSS, the AMI corpus introduces severe challenges due to spontaneous conversational dynamics and real-room acoustics. AMI contains frequent short speech bursts that complicate speaker tracking. Furthermore, real-world reverberation and background noise degrade speaker separation quality. Additionally, AMI meetings average 33 minutes in duration, posing significant challenges for online diarization systems to maintain speaker identity over extended periods. 

\autoref{tab:AMIDer} reports the performances of the proposed and comparison systems. Among offline methods, AED-EEND-EE with Conformer~\cite{chentaslp2024} achieves the lowest \gls{der} of \SI{13.00}{\percent}. Among online methods, our speaker tracking system obtains the best \gls{der} of 18.43\%, outperforming LS-EEND~\cite{liang2025ls} and other online alternatives. These results demonstrate the robustness of our speaker tracking framework in realistic meeting scenarios.
\input{Tables/AMI.tex}

\subsection{Ablation study}
\label{subsec:ablation_analysis}
To isolate the effects of individual system components, we conduct an ablation study analyzing speaker models, \gls{vad} models, segment sizes, and segment overlaps. \autoref{tab:ablation} summarizes the tracking performance in \gls{der} and \gls{cpwer} evaluated on the seven-channel LibriCSS dataset.

Comparing speaker models, ReDimNet yields the lowest \gls{der} ({3.38\%}) and \gls{cpwer} ({5.32\%}), outperforming the alternatives of ReDimNet2, Wespeaker, NeMo, and SpeechBrain. The performance differences indicate that discriminative embeddings play a constructive role in speaker tracking.

Evaluating \gls{vad} models indicates that Pyannote provides the best performance, achieving the lowest \gls{der} and \gls{cpwer}. Silero yields slightly higher \gls{der}. Substituting the \gls{vad} component by WebRTC significantly increases diarization errors.

Segment length affects speaker tracking. A segment size of \SI{4.0}{\second} yields the best \gls{der} (3.38\%). Enlarging the segment size  degrades performance (e.g., 5.35\% \gls{der} for \SI{8.0}{\second}). Longer segments increase the number of speakers within a segment, causing the total number of speakers to potentially surpass the capacity of the speaker separation model ($\nspkMax = 4$). Conversely, shrinking the segment size to \SI{2.0}{\second} also degrades performance ({6.75\%} \gls{der}), due to insufficient contextual information for stitching and speaker embedding.

Increasing segment overlap degrades tracking performance. A \SI{25}{\percent} overlap seems to provide the sufficient context for overlap-based stitching. Larger overlaps compel the system to perform more sequential grouping operations, increasing \gls{der}. For example, expanding segment overlap from \SI{25}{\percent} to \SI{50}{\percent} raises \gls{der} from {3.38\%} to {4.41\%}.

\input{Tables/ablation.tex}

\autoref{tab:vertical_comparison} provides a detailed breakdown of \gls{der} components across several configurations, revealing the impacts of error types on downstream \gls{cpwer}. The proposed system with different configurations exhibits relatively balanced error distributions, with \gls{mi} contributions ranging from {1.97\%} to {2.03\%}. Compared to the single-channel system, multi-channel system substantially reduces \gls{cf} errors (from 1.63\% to 0.48\%). The corresponding results from two SSND versions, one based on TF-GridNet for separation as reported in \autoref{tab:diarization_results} and another based on SpatialNet for separation as reported in \autoref{tab:LibriCSS_cpWER}. The second \gls{ssnd} configuration produces higher \gls{fa} errors (\SI{6.98}{\percent}), but the lower \gls{cpwer} of \SI{5.13}{\percent} despite elevated \gls{der} (\SI{7.62}{\percent}). This seemingly contradictory pattern underscores that \gls{fa} errors during silence periods incur minimal ASR penalty. The oracle configuration achieves the lowest \gls{cf} (\SI{0.48}{\percent}) and the lowest \gls{cpwer} (\SI{3.66}{\percent}), highlighting that confusion between speakers remains a bottleneck. These ablation results demonstrate that minimizing \gls{mi} and \gls{cf} is essential for speaker tracking systems employed in downstream \gls{asr} pipelines, whereas \gls{fa} errors are secondary.

Compared to \gls{ssnd}~\cite{taherian2024multi}, our system achieves both high diarization and high \gls{asr} results within the same model. In contrast, \gls{ssnd} utilizes separate models for optimal diarization and \gls{asr}. Additionally, we employ a sophisticated \gls{vad} module to lower \gls{fa} errors. 

\input{Tables/DER_breakdown.tex}

\begin{figure}[th]
    \centering
    \def\svgwidth{0.99\linewidth}
    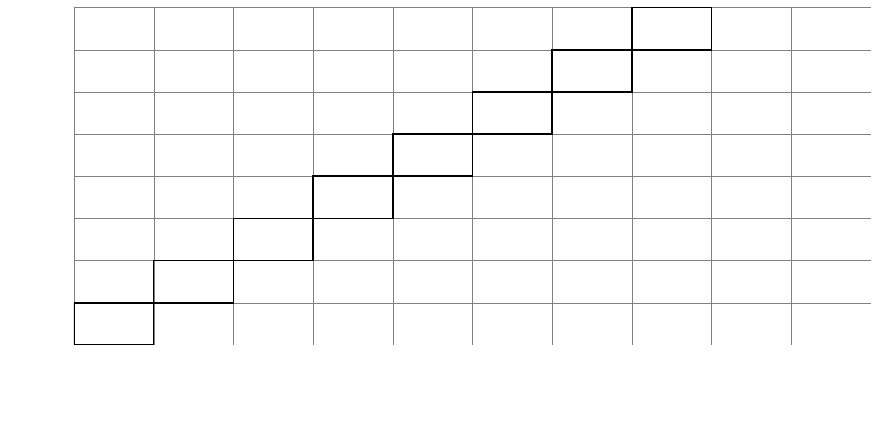
    \caption{Speaker counting accuracy of the speaker tracking model evaluated on the section-wise LibriCSS data.}
    \label{fig:speaker_counting}
\end{figure}

\autoref{fig:speaker_counting} illustrates the speaker count performance of the proposed speaker tracking method on the section-wise LibriCSS dataset. For each section, we run the proposed speaker tracking method and count the number of predicted tracks, which is compared against the actual number of speakers in that section. The results in \autoref{fig:speaker_counting} are averaged over all sections. The system achieves very high accuracy for mixture audio with few speakers, correctly estimating the speaker count \SI{100}{\percent} and \SI{97.06}{\percent} for one and two actual speakers, respectively. As the number of actual speakers increases, the accuracy declines, and the system tends to overestimate the speaker count. For example, with seven actual speakers, the system correctly identifies seven speakers in \SI{62.69}{\percent} of the segments and estimates eight speakers in \SI{28.36}{\percent} of segments. This overestimation is primarily attributed to track fragmentation and \gls{fa} errors, which breaks the same speaker into multiple tracks. On the other hand, underestimation is infrequent. These results suggest that the proposed framework robustly initiates new speaker tracks, but with a tendency to over-initiate.

%% file: Tables/diarization.tex
\begin{table*}[t]
  \centering
  \caption{DER results (in \%) of the proposed and comparison diarization systems on LibriCSS session-wise evaluation.}
  \begin{adjustbox}{width=0.70\textwidth}
  \begin{threeparttable}
    \begin{tabular}{clcc ccccc}
      \toprule
      \multirow{2}{*}{Separation method} & \multirow{2}{*}{Diarization method} & \multicolumn{6}{c}{Overlap ratio} & \multirow{2}{*}{All} \\
      \cmidrule(r){3-8}
      & & 0L & 0S & 10\% & 20\% & 30\% & 40\% \\
      \midrule
      Mask-based MVDR~\cite{chen2020continuous}  & {X-vector+SC}~\cite{rajSLTIntegration}                          & 13.42 & 11.49 & 11.63 & 14.22 & 16.98 & 16.18 & 14.18 \\
      Multi-Frame MCWF~\cite{wang2021sequential} & {X-vector+SC}~\cite{rajSLTIntegration}                          & 11.80 & 10.20 & 13.84 & 15.76 & 18.69 & 17.33 & 15.04 \\
      MISO-BF-MISO~\cite{wang2021multi}          & DOA-based~\cite{wang2022localization}                           & 10.69 & 11.95 & 11.25 & 12.22 & 13.04 & 14.31 & 12.36 \\
      MIMO-BF-MISO~\cite{taherianICASSP24}          & {X-vector+SC}~\cite{rajSLTIntegration}                          & 10.34  & 10.4  & 8.97  & 9.52  & 11.66 & 9.54  & 9.9  \\

      -                                          & MC-EEND~\cite{taherian2024multi}                          & 6.12  & 4.94  & 3.36  & 4.09  & 4.88  & 5.05  & 4.68  \\
      \midrule
      -                                          & {X-vector+SC}~\cite{rajSLTIntegration} & 10.25 & 9.29  & 14.04 & 18.76 & 23.82 & 27.43 & 18.19 \\
      -                                          & AHC~\cite{rajSLTIntegration}                                      & 16.1 & 12.0 & 16.9  & 23.6  & 28.3  & 33.2  & 22.6  \\
      -                                          & VBx~\cite{rajSLTIntegration}                                      & 14.6 & 11.1  & 14.3  & 21.5  & 25.4  & 31.2  & 20.5  \\
      -                                          & SC~\cite{rajSLTIntegration}                                     & 10.9 & 9.5   & 13.9  & 18.9  & 23.7  & 27.4  & 18.3  \\
      -                                          & SC-EEND~\cite{taherian2024multi}                                & 23.48 & 23.2  & 29.55 & 27.1  & 34.94 & 39.22 & 30.36 \\
      -                                          & RPN\tnote{*}~\cite{rajSLTIntegration}                           & 4.5   & 9.1   & 8.3   & 6.7   & 11.6  & 14.2  & 9.5   \\
      -                                          & TS-VAD\tnote{*}~\cite{rajSLTIntegration}                        & 6.0   & 4.6   & 6.6   & 7.3   & 10.3  & 9.5   & 7.6   \\
      LlaSep~\cite{lanzendoerfer26b_interspeech} & -                                         & 18.57 & 21.91 & 24.79 & 26.85 & 20.96 & 27.52 & 23.43  \\
      \midrule
      TF-CrossNet (1-ch)                         & Sequential grouping                                               & {3.43} & {3.57} & {4.63} & {4.20} & {6.05} & {5.25} & {4.63} \\

      TF-CrossNet (7-ch)                         & Sequential grouping                                          & {3.22} & {3.50} & {2.86} & {3.22} & {3.95} & {3.35} & {3.38} \\

      \midrule
      Oracle                                      & Sequential grouping                                            & 3.12 & 2.83 & 2.69 & 2.63 & 3.68 & 2.74 & 2.98 \\

      \bottomrule
    \end{tabular}
    \begin{tablenotes}
      \item[*] Assumes that the oracle number of speakers is known.
    \end{tablenotes}
  \end{threeparttable}
  \end{adjustbox}
  \label{tab:diarization_results}
\end{table*}

%% file: Tables/LibriCSS_WER.tex
\begin{table*}[t]
  \centering
  \renewcommand{\arraystretch}{1.25}
  \caption{cpWER results (in \%) for different separation and diarization methods on LibriCSS session-wise evaluation.}
  \begin{adjustbox}{width=0.8\textwidth}
    \begin{tabular}{l c c c cccccccc}
      \toprule

      \multirow{2}{*}{Separation method} & \multirow{2}{*}{Diarization method} & \multirow{2}{*}{ASR} & \multirow{2}{*}{DER (\%)} & \multicolumn{6}{c}{Overlap ratio} & \multirow{2}{*}{All} \\
      \cmidrule(r){5-10}
      & & & & 0L & 0S & 10\% & 20\% & 30\% & 40\% \\
      \midrule
      Unprocessed                                      & X-vector + SC        & E2E     & 26.76 & 12.2  & 13.95 & 20.12 & 29.64 & 35.06 & 41.81 & 27.01 \\
      \midrule
      Mask-based MVDR~\cite{chen2020continuous}        & X-vector + SC        & E2E     & 14.18 & 13.07 & 8.78  & 10.51 & 15.37 & 17.54 & 17.63 & 14.13 \\
      
      MISO-BF-MISO~\cite{wang2021multi}         & X-vector + SC        & E2E     & 8.33  & 9.24  & 10.84 & 11.05 & 12.27 & 16.39 & 17.42 & 12.87 \\
      MISO-BF-MISO~\cite{wang2021multi}         & DOA-based            & E2E     & 12.36 & 9.03  & 10.18 & 10.89 & 13.6  & 16.14 & 18.05 & 12.98 \\

      MIMO-BF-MISO~\cite{taherianICASSP24}             & X-vector + SC        & E2E     & 9.90  & 8.05  & 7.05  & 7.31  & 8.48  & 10.81 & 10.03 & 8.76  \\

      TF-GridNet~\cite{von2024meeting}  & ASR-based& E2E-SSL & 9.4 & --& --& --&-- & --& --& 7.2 \\

      DCF-DS~\cite{niu2025dcf} & Joint & E2E &-- & --& --&-- &-- & --&-- & 6.30 \\
      SSND~\cite{taherian2024multi}       & MC-EEND & E2E     & 7.62  & 3.52  & 5.56  & 3.98  & 4.76  & 5.58  & 6.55  & 5.13  \\
      SSND~\cite{taherian2024multi}       & MC-EEND & E2E-SSL & 7.62  & 1.97  & 3.67  & 2.46  & 3.06  & 3.52  & 4.07  & 3.22  \\
      
      TMS~\cite{liu2026tms} & -- & E2E-SSL & -- & 1.98 & 3.10 & 2.05 & 2.55 & 2.80 & 3.40 & 2.64 \\
      \midrule

      TF-CrossNet (1-ch)~\cite{kalkhorani2024crossnet}       & Sequential grouping     & E2E & {4.63} & {5.62} & {6.00} & {8.30} & {8.53} & {12.62} & {12.07} & {9.25} \\
      TF-CrossNet (1-ch)~\cite{kalkhorani2024crossnet}       & Sequential grouping     & E2E-SSL & {4.63} & {4.51} & {4.82} & {6.25} & {6.36} & {10.14} & {9.47} & {7.20} \\

      TF-CrossNet (7-ch)~\cite{kalkhorani2024crossnet}       & Sequential grouping     & E2E     & {3.38} & {4.52} & {4.74} & {4.16} & {5.26} & {6.58} & {6.03} & {5.32} \\

      TF-CrossNet (7-ch)~\cite{kalkhorani2024crossnet}       & Sequential grouping     & E2E-SSL   & {3.38} & {4.32} & {3.22} & {3.00} & {4.20} & {5.53} & {5.49} & {4.39} \\

      \midrule
      Oracle & Sequential grouping & E2E & 2.98 & 3.39 & 2.75 & 3.10 & 3.01 & 5.78 & 3.53 & 3.66 \\
      Oracle & Sequential grouping & E2E-SSL & 2.98 & 2.78 & 2.21 & 2.90 & 2.62 & 4.72 & 3.28 & 3.15  \\


      \bottomrule
    \end{tabular}
  \end{adjustbox}
  \label{tab:LibriCSS_cpWER}
\end{table*}

%% file: Tables/AMI.tex
\begin{table}[t]
    \caption{\Gls{der} results (in \%) of the proposed and comparison diarization systems on AMI evaluation set.}
    \label{tab:AMIDer}
    \centering
    \begin{adjustbox}{width=0.4\textwidth}
        \begin{tabular}{lc}
            \toprule
            Methods                                                               & DER (\%)    \\
            \midrule
            \textbf{Offline} \\
            \quad X-vector AHC + VB + OVL \cite{Horiguchi2021end, Garcia2020odyssey} & 28.15   \\
            \quad SA-EEND \cite{fujita2019end, horiguchi2022encoder}              & 27.70   \\
            \quad Transcribe-to-Diarize \cite{kanda2022transcibe}                 & 24.43   \\
            \quad Horiguchi~\etal~\cite{horiguchi26_interspeech}       & 23.62   \\
            \quad Multi-Class Spec-Clustering \cite{raj2021multiclass}            & 23.60   \\
            \quad CmpEm + Overlap Detector \cite{li2021compositional}             & 22.93   \\
            \quad EENDA-EDA \cite{horiguchi20_interspeech, horiguchi2022encoder}  & 21.56   \\
            \quad ANSD-MA-MSE \cite{he2023ansd}                                    & 16.95   \\
            \quad AED-EEND-EE + Conformer \cite{chentaslp2024}                    & 13.00   \\
            \midrule

            \textbf{Online} \\
            \quad Overlap-aware speaker embeddings \cite{coria2021overlap}        & 30.4    \\
            \quad SSep AMI + VAD E2E \cite{gruttadauria2024online}                & 27.2    \\
            \quad {LS-EEND~\cite{liang2025ls}}                                    & 20.76   \\
            \quad Speaker tracking & {18.43}   \\

            \bottomrule                                                                                                  
        \end{tabular}
		\end{adjustbox}
	
\end{table}

%% file: Tables/ablation.tex
\begin{table}[t]
\centering
\caption{Ablation study evaluating the impacts of speaker embedding models, \gls{vad}, segment lengths and overlaps on \gls{der} and \gls{cpwer}.}
\label{tab:ablation}
\begin{adjustbox}{width=0.49\textwidth}
\begin{tabular}{llcc S[table-format=2.2, detect-weight=true] S[table-format=2.2, detect-weight=true]}
\toprule
$\modelSPK$ & $\modelVAD$ & $\seqLen$ (s) & $\seqLenOvlap/\seqLen$ & {DER (\%)} & {cpWER (\%)}  \\
\midrule
ReDimNet~\cite{yakovlev2024redimenet} & \multirow{5}{*}{Pyannote~\cite{bredin2020pyannote}} & \multirow{5}{*}{4.0} & \multirow{5}{*}{0.25} & {{3.38}} & {{5.32}} \\
ReDimNet2~\cite{yakovlev2026redimnet2} & & & & {{3.66}} & {{5.56}} \\
Wespeaker~\cite{wang2023wespeaker} & & & & {{4.18}} & {{6.69}} \\
NeMo~\cite{kuchaiev2019nemo} & & & & {{5.10}} & {{8.73}} \\
SpeechBrain~\cite{ravanelli2021speechbrain} & & & & {{5.77}} & {{9.33}} \\
\midrule
\multirow{2}{*}{ReDimNet~\cite{yakovlev2024redimenet}} & Silero & \multirow{2}{*}{4.0} & \multirow{2}{*}{0.25} & {{4.07}} & {{5.50}} \\
& WebRTC & & & {{6.83}} & {{5.71}} \\
\midrule
\multirow{5}{*}{ReDimNet~\cite{yakovlev2024redimenet}} & \multirow{5}{*}{Pyannote~\cite{bredin2020pyannote}} & 2.0 & \multirow{5}{*}{0.25} & {{6.75}} & {{10.11}} \\
& & 3.0 & & {{4.39}} & {{6.90}} \\
& & 5.0 & & {{3.69}} & {{5.72}} \\
& & 6.0 & & {{4.14}} & {{6.34}} \\
& & 8.0 & & {{5.35}} & {{7.52}} \\
\midrule
\multirow{2}{*}{ReDimNet~\cite{yakovlev2024redimenet}} & \multirow{2}{*}{Pyannote~\cite{bredin2020pyannote}} & \multirow{2}{*}{4.0} & 0.50 & {{4.41}} & {{6.76}} \\
& & & 0.75 & {{7.99}} & {{11.63}} \\
\bottomrule
\end{tabular}
\end{adjustbox}
\end{table}

%% file: Tables/DER_breakdown.tex
\begin{table}[ht]
\centering
\caption{Breakdown of \gls{der} components (\gls{mi}, \gls{fa}, \gls{cf}) across system configurations and acoustic conditions on LibriCSS dataset.}
\label{tab:vertical_comparison}
\begin{adjustbox}{width=0.49\textwidth}
    \begin{tabular}{lccccc}
        \toprule
        System  & \gls{der} & \gls{mi} & \gls{fa} & \gls{cf} & \gls{cpwer} \\

        \midrule
        SSND (SpatialNet) & 7.62 & 0.53 & 6.98 & 0.11 & 5.13 \\
        SSND (TF-GridNet) & 4.68 & 1.86 & 2.16 & 0.66 & 6.84 \\

        \midrule
        \textbf{Speaker tracking}  \\
        \quad 1-ch, $\seqLen=\SI{4}{\second}$, $ \seqLenOvlap / \seqLen=0.25$ & {4.63} & {1.97} & {1.03} & {1.63} & {9.25} \\
        \quad 7-ch, $\seqLen=\SI{5}{\second}$, $ \seqLenOvlap / \seqLen=0.25$ & {3.69} & {2.03} & {1.04} & {0.62} & {5.72} \\

        \quad 7-ch, $\seqLen=\SI{4}{\second}$, $ \seqLenOvlap / \seqLen=0.25$ & {3.38} & {2.02} & {0.88} & {0.48} & {5.32} \\

        \quad Oracle, $\seqLen=\SI{4}{\second}$, $ \seqLenOvlap / \seqLen=0.25$ & 2.98 & 1.33 & 1.17 & 0.48 & 3.66 \\
        \bottomrule
    \end{tabular}
\end{adjustbox}
\end{table}

%% file: Figures/Speaker_counting.pdf_tex
\begingroup%
  \makeatletter%
  \providecommand\color[2][]{%
    \errmessage{(Inkscape) Color is used for the text in Inkscape, but the package 'color.sty' is not loaded}%
    \renewcommand\color[2][]{}%
  }%
  \providecommand\transparent[1]{%
    \errmessage{(Inkscape) Transparency is used (non-zero) for the text in Inkscape, but the package 'transparent.sty' is not loaded}%
    \renewcommand\transparent[1]{}%
  }%
  \providecommand\rotatebox[2]{#2}%
  \newcommand*\fsize{\dimexpr\f@size pt\relax}%
  \newcommand*\lineheight[1]{\fontsize{\fsize}{#1\fsize}\selectfont}%
  \ifx\svgwidth\undefined%
    \setlength{\unitlength}{421.62545776bp}%
    \ifx\svgscale\undefined%
      \relax%
    \else%
      \setlength{\unitlength}{\unitlength * \real{\svgscale}}%
    \fi%
  \else%
    \setlength{\unitlength}{\svgwidth}%
  \fi%
  \global\let\svgwidth\undefined%
  \global\let\svgscale\undefined%
  \makeatother%
  \footnotesize
  \begin{picture}(1,0.48777742)%
    \lineheight{1}%
    \setlength\tabcolsep{0pt}%
    \put(0,0){\includegraphics[width=\unitlength,page=1]{Figures/Speaker_counting.pdf}}%
    \put(0.12954424,0.06165796){\makebox(0,0)[t]{\lineheight{1.25}\smash{\begin{tabular}[t]{c}1\end{tabular}}}}%
    \put(0.2202724,0.06165796){\makebox(0,0)[t]{\lineheight{1.25}\smash{\begin{tabular}[t]{c}2\end{tabular}}}}%
    \put(0.31100055,0.06165796){\makebox(0,0)[t]{\lineheight{1.25}\smash{\begin{tabular}[t]{c}3\end{tabular}}}}%
    \put(0.40172869,0.06165796){\makebox(0,0)[t]{\lineheight{1.25}\smash{\begin{tabular}[t]{c}4\end{tabular}}}}%
    \put(0.49245687,0.06165796){\makebox(0,0)[t]{\lineheight{1.25}\smash{\begin{tabular}[t]{c}5\end{tabular}}}}%
    \put(0.58318502,0.06165796){\makebox(0,0)[t]{\lineheight{1.25}\smash{\begin{tabular}[t]{c}6\end{tabular}}}}%
    \put(0.67391316,0.06165796){\makebox(0,0)[t]{\lineheight{1.25}\smash{\begin{tabular}[t]{c}7\end{tabular}}}}%
    \put(0.76464131,0.06165796){\makebox(0,0)[t]{\lineheight{1.25}\smash{\begin{tabular}[t]{c}8\end{tabular}}}}%
    \put(0.85536946,0.06165796){\makebox(0,0)[t]{\lineheight{1.25}\smash{\begin{tabular}[t]{c}9\end{tabular}}}}%
    \put(0.94609767,0.06165796){\makebox(0,0)[t]{\lineheight{1.25}\smash{\begin{tabular}[t]{c}10\end{tabular}}}}%
    \put(0.53782095,0.01643047){\makebox(0,0)[t]{\lineheight{1.25}\smash{\begin{tabular}[t]{c}Estimated number of speakers\end{tabular}}}}%
    \put(0.07587896,0.11003029){\makebox(0,0)[rt]{\lineheight{1.25}\smash{\begin{tabular}[t]{r}1\end{tabular}}}}%
    \put(0.07587896,0.15803644){\makebox(0,0)[rt]{\lineheight{1.25}\smash{\begin{tabular}[t]{r}2\end{tabular}}}}%
    \put(0.07587896,0.20604263){\makebox(0,0)[rt]{\lineheight{1.25}\smash{\begin{tabular}[t]{r}3\end{tabular}}}}%
    \put(0.07587896,0.2540488){\makebox(0,0)[rt]{\lineheight{1.25}\smash{\begin{tabular}[t]{r}4\end{tabular}}}}%
    \put(0.07587896,0.30205497){\makebox(0,0)[rt]{\lineheight{1.25}\smash{\begin{tabular}[t]{r}5\end{tabular}}}}%
    \put(0.07587896,0.35006115){\makebox(0,0)[rt]{\lineheight{1.25}\smash{\begin{tabular}[t]{r}6\end{tabular}}}}%
    \put(0.07587896,0.39806732){\makebox(0,0)[rt]{\lineheight{1.25}\smash{\begin{tabular}[t]{r}7\end{tabular}}}}%
    \put(0.07587896,0.44607349){\makebox(0,0)[rt]{\lineheight{1.25}\smash{\begin{tabular}[t]{r}8\end{tabular}}}}%
    \put(0.03737322,0.28721434){\rotatebox{90}{\makebox(0,0)[t]{\lineheight{1.25}\smash{\begin{tabular}[t]{c}Actual number of speakers\end{tabular}}}}}%
    \put(0,0){\includegraphics[width=\unitlength,page=2]{Figures/Speaker_counting.pdf}}%
    \put(0.12954424,0.11133919){\color[rgb]{0.14901961,0.14901961,0.14901961}\makebox(0,0)[t]{\lineheight{1.25}\smash{\begin{tabular}[t]{c}100\%\end{tabular}}}}%
    \put(0.2202724,0.15934538){\color[rgb]{0.14901961,0.14901961,0.14901961}\makebox(0,0)[t]{\lineheight{1.25}\smash{\begin{tabular}[t]{c}97.1\%\end{tabular}}}}%
    \put(0.31100055,0.15934538){\color[rgb]{0.14901961,0.14901961,0.14901961}\makebox(0,0)[t]{\lineheight{1.25}\smash{\begin{tabular}[t]{c}2.9\%\end{tabular}}}}%
    \put(0.31100055,0.20735155){\color[rgb]{0.14901961,0.14901961,0.14901961}\makebox(0,0)[t]{\lineheight{1.25}\smash{\begin{tabular}[t]{c}85.4\%\end{tabular}}}}%
    \put(0.40172869,0.20735155){\color[rgb]{0.14901961,0.14901961,0.14901961}\makebox(0,0)[t]{\lineheight{1.25}\smash{\begin{tabular}[t]{c}13.5\%\end{tabular}}}}%
    \put(0.49245687,0.20735155){\color[rgb]{0.14901961,0.14901961,0.14901961}\makebox(0,0)[t]{\lineheight{1.25}\smash{\begin{tabular}[t]{c}1.1\%\end{tabular}}}}%
    \put(0.40172869,0.25535772){\color[rgb]{0.14901961,0.14901961,0.14901961}\makebox(0,0)[t]{\lineheight{1.25}\smash{\begin{tabular}[t]{c}84.2\%\end{tabular}}}}%
    \put(0.49245687,0.25535772){\color[rgb]{0.14901961,0.14901961,0.14901961}\makebox(0,0)[t]{\lineheight{1.25}\smash{\begin{tabular}[t]{c}14.1\%\end{tabular}}}}%
    \put(0.58318502,0.25535772){\color[rgb]{0.14901961,0.14901961,0.14901961}\makebox(0,0)[t]{\lineheight{1.25}\smash{\begin{tabular}[t]{c}1.7\%\end{tabular}}}}%
    \put(0.40172869,0.3033639){\color[rgb]{0.14901961,0.14901961,0.14901961}\makebox(0,0)[t]{\lineheight{1.25}\smash{\begin{tabular}[t]{c}1.1\%\end{tabular}}}}%
    \put(0.49245687,0.3033639){\color[rgb]{0.14901961,0.14901961,0.14901961}\makebox(0,0)[t]{\lineheight{1.25}\smash{\begin{tabular}[t]{c}80.7\%\end{tabular}}}}%
    \put(0.58318502,0.3033639){\color[rgb]{0.14901961,0.14901961,0.14901961}\makebox(0,0)[t]{\lineheight{1.25}\smash{\begin{tabular}[t]{c}17.1\%\end{tabular}}}}%
    \put(0.67391316,0.3033639){\color[rgb]{0.14901961,0.14901961,0.14901961}\makebox(0,0)[t]{\lineheight{1.25}\smash{\begin{tabular}[t]{c}1.1\%\end{tabular}}}}%
    \put(0.49245687,0.35137006){\color[rgb]{0.14901961,0.14901961,0.14901961}\makebox(0,0)[t]{\lineheight{1.25}\smash{\begin{tabular}[t]{c}2.9\%\end{tabular}}}}%
    \put(0.58318502,0.35137006){\color[rgb]{0.14901961,0.14901961,0.14901961}\makebox(0,0)[t]{\lineheight{1.25}\smash{\begin{tabular}[t]{c}76.3\%\end{tabular}}}}%
    \put(0.67391316,0.35137006){\color[rgb]{0.14901961,0.14901961,0.14901961}\makebox(0,0)[t]{\lineheight{1.25}\smash{\begin{tabular}[t]{c}18.0\%\end{tabular}}}}%
    \put(0.76464131,0.35137006){\color[rgb]{0.14901961,0.14901961,0.14901961}\makebox(0,0)[t]{\lineheight{1.25}\smash{\begin{tabular}[t]{c}2.2\%\end{tabular}}}}%
    \put(0.85536946,0.35137006){\color[rgb]{0.14901961,0.14901961,0.14901961}\makebox(0,0)[t]{\lineheight{1.25}\smash{\begin{tabular}[t]{c}0.7\%\end{tabular}}}}%
    \put(0.49245687,0.39937624){\color[rgb]{0.14901961,0.14901961,0.14901961}\makebox(0,0)[t]{\lineheight{1.25}\smash{\begin{tabular}[t]{c}1.5\%\end{tabular}}}}%
    \put(0.58318502,0.39937624){\color[rgb]{0.14901961,0.14901961,0.14901961}\makebox(0,0)[t]{\lineheight{1.25}\smash{\begin{tabular}[t]{c}4.5\%\end{tabular}}}}%
    \put(0.67391316,0.39937624){\color[rgb]{0.14901961,0.14901961,0.14901961}\makebox(0,0)[t]{\lineheight{1.25}\smash{\begin{tabular}[t]{c}62.7\%\end{tabular}}}}%
    \put(0.76464131,0.39937624){\color[rgb]{0.14901961,0.14901961,0.14901961}\makebox(0,0)[t]{\lineheight{1.25}\smash{\begin{tabular}[t]{c}28.4\%\end{tabular}}}}%
    \put(0.85536946,0.39937624){\color[rgb]{0.14901961,0.14901961,0.14901961}\makebox(0,0)[t]{\lineheight{1.25}\smash{\begin{tabular}[t]{c}3.0\%\end{tabular}}}}%
    \put(0.76464131,0.44738241){\color[rgb]{0.14901961,0.14901961,0.14901961}\makebox(0,0)[t]{\lineheight{1.25}\smash{\begin{tabular}[t]{c}76.2\%\end{tabular}}}}%
    \put(0.85536946,0.44738241){\color[rgb]{0.14901961,0.14901961,0.14901961}\makebox(0,0)[t]{\lineheight{1.25}\smash{\begin{tabular}[t]{c}19.0\%\end{tabular}}}}%
    \put(0.94609767,0.44738241){\color[rgb]{0.14901961,0.14901961,0.14901961}\makebox(0,0)[t]{\lineheight{1.25}\smash{\begin{tabular}[t]{c}4.8\%\end{tabular}}}}%
  \end{picture}%
\endgroup%

%% file: Sections/06-Conclusions.tex
\vspace{-0.2cm}
\section{Conclusions}
\label{sec:conclusion}
We have proposed a segment-online system for speaker tracking. The system performs speaker separation for simultaneous grouping, yielding the estimates of overlapped speakers. These estimates are further organized sequentially into coherent speakers, each corresponding to a distinct talker. Sequential organization is performed in two stages. The first stage employs overlap-based stitching for continuous organization and the second stage uses working memory based on speaker verification for discontinuous organization. Experimental results show our system advances the state-of-the-art in speaker diarization on LibriCSS. On AMI, we achieve the best diarization results in the segment-online setting. Our system also obtains better speaker-attributed ASR results than other methods, and competitive performance to offline methods. For future work, we plan to explore frame-online processing to enhance real-time processing capability. We also plan to address track fragmentation and false-alarm errors that degrade speaker count accuracy as revealed in \autoref{fig:speaker_counting}.

%% file: Sections/07-Acknowledgement.tex
\section{Acknowledgment}

The first author was supported in part by a Meta contract to Ohio State University, a National Science Foundation grant (ECCS-2125074), the Ohio Supercomputer Center (OSC) and the NCSA Delta Supercomputer Center (OCI 2005572).